\documentclass[aps,prx,amsmath,floats,floatfix,twocolumn,
  superscriptaddress,nofootinbib,showpacs]{revtex4-1}

\pdfoutput=1 
\usepackage{diagbox}
\usepackage{multirow}
\usepackage{braket}
\usepackage{amsfonts}
\usepackage{amsmath}
\usepackage{amssymb}
\usepackage{amsthm}
\usepackage{bm}
\usepackage{graphicx}
\usepackage{graphics}
\usepackage{latexsym}
\usepackage{rotating}
\usepackage[dvipsnames]{xcolor}
\usepackage{mathrsfs}
\usepackage{microtype}
\usepackage{verbatim}
\usepackage{url}

\usepackage[caption=false]{subfig}
\usepackage[normalem]{ulem}

\usepackage[breaklinks=true]{hyperref}
\hypersetup{
    colorlinks=true,
    linkcolor=NavyBlue,
    filecolor=Magenta,
    urlcolor=NavyBlue,
    citecolor=NavyBlue
}

\usepackage{yfonts}
\usepackage{float}
\usepackage{xspace} 
\usepackage{mathrsfs}
\usepackage{siunitx}
\usepackage{array}

\allowdisplaybreaks[4]

\newcommand{\be}{\begin{equation}}
\newcommand{\ee}{\end{equation}}

\newcommand{\ba}{\begin{align}}
\newcommand{\ea}{\end{align}}

\makeatletter
\newcommand*{\rom}[1]{\expandafter\@slowromancap\romannumeral #1@}
\makeatother

\AtBeginDocument{%
    \newwrite\bibnotes
    \def\bibnotesext{Notes.bib}
    \immediate\openout\bibnotes=\jobname\bibnotesext
    \immediate\write\bibnotes{@CONTROL{REVTEX41Control}}
    \immediate\write\bibnotes{@CONTROL{%
    apsrev41Control,author="08",editor="1",pages="1",title="0",year="1"}}
    \if@filesw
    \immediate\write\@auxout{\string\citation{apsrev41Control}}%
    \fi
}

\allowdisplaybreaks

\begin{document}


\title{Quasinormal mode frequencies and gravitational perturbations of spinning black holes in modified gravity through METRICS: \\ The dynamical Chern-Simons gravity case }

\author{Adrian Ka-Wai Chung}
\email{akwchung@illinois.edu}
\affiliation{Illinois Center for Advanced Studies of the Universe and Department of Physics, University of Illinois Urbana-Champaign, Urbana, Illinois 61801, USA}

\author{Kelvin Ka-Ho Lam}
\email{khlam4@illinois.edu}
\affiliation{Illinois Center for Advanced Studies of the Universe and Department of Physics, University of Illinois Urbana-Champaign, Urbana, Illinois 61801, USA}

\author{Nicol\'as Yunes}
\affiliation{Illinois Center for Advanced Studies of the Universe and Department of Physics, University of Illinois Urbana-Champaign, Urbana, Illinois 61801, USA}

\date{\today}

\begin{abstract}

We present the first precise calculations of the gravitational quasinormal-mode (QNM) frequencies for spinning black holes with dimensionless angular momenta $J/M^2 := a \lesssim 0.75$ in dynamical Chern-Simons gravity. Using the \textit{Metric pErTuRbations wIth speCtral methodS} (METRICS) framework, we compute the QNM frequencies of both axial and polar metric perturbations, focusing on the $nl m = 022$, $033$, and $032$ modes.
The METRICS frequencies for the 022 mode achieve numerical uncertainties $\lesssim 10^{-4}$ when $0 \leq a \leq 0.5$ and $\lesssim 10^{-3}$ for $0.5 \leq a \leq 0.75$, without decoupling or simplifying the linearized field equations. 
We also derive optimal fitting polynomials to enable efficient and accurate evaluations of the leading-order frequency shifts in these modes. 
The METRICS frequencies and fitting expressions are a robust and indispensable step toward enabling gravitational-wave ringdown tests of dynamical Chern-Simons gravity.

\end{abstract}

\maketitle


\section{Introduction}
\label{sec:intro}

The validity of general relativity (GR) is a central aspect of fundamental physics research in gravitation. 
GR has passed all experimental and observational tests~\cite{Will2014,Stairs2003,Wex:2020ald,Yunes:2013dva,Will2014,Yagi:2016jml,Berti:2018cxi,Nair:2019iur,Berti:2018vdi, LIGO_07, LIGO_11, LIGOScientific:2021sio, Perkins:2022fhr, Yunes:2013dva}, making it the most accurate gravity theory to date. 
However, GR presents some ``anomalies'' that have prompted many physicists to consider extensions.  
Theoretically, GR predicts that the formation of a spacetime singularity is the ultimate consequence of gravitational collapse~\cite{PhysRevLett.14.57, Hawking:1970zqf}, at least at the classical level. 
Yet, GR is unable to elucidate the nature of a spacetime singularity, where the theory itself breaks down and loses predictability.
Observationally, GR cannot explain some astrophysical phenomena, such as galaxy rotation curve~\cite{rotation_curve_01, rotation_curve_02}, and the acceleration of the Universe at late times~\cite{late_time_acceleration_01, late_time_acceleration_02} without invoking dark matter or dark energy, whose nature remains unknown. 
GR also fails to explain the matter-antimatter asymmetry without additional parity-violating physics in the early universe~\cite{Sakharov:1967dj,Petraki:2013wwa, Gell-Mann:1991kdm, Alexander:2004us}.
These unresolved anomalies have motivated various modifications to GR. 
Identifying the correct modification or ruling out extensions requires observational or experimental tests. 

Binary black hole (BBH) coalescence serves as an ideal test bed for GR (and its modifications) because its gravitational-wave (GW) signals encode information about the strong-field dynamics governing these system.
In general, BBH coalescence is composed of the following three stages. 
The first is the inspiral stage, where the two BHs orbit each other, emitting GWs. 
These waves carry away energy and angular momentum, causing the orbit to shrink and leading to an increase in both the amplitude and frequency of the emitted GWs. 
The second stage is the merger, during which the two BHs collide and merge into a single but distorted remnant BHs, producing GWs of the largest amplitude through the entire process. 
The third and final stage is the ringdown, during which the remnant BH settles to a stationary state by emitting GWs. After an initial nonlinear phase, these GWs consist mostly\footnote{Strictly speaking, power-law tails will also emerge during the sufficiently late part of the ringdown stage. 
However, as power-law tails may not be detectable, we shall omit them from our discussion in this paper.} of exponentially decaying sinusoids. 
The real frequency and the decay lifetime of these sinusoidal waves are collectively called ``quasinormal-mode (QNM) frequencies.''

Compact objects (including BHs) still exist in beyond Einstein theories, but their spacetime is different from that in GR. 
Moreover, the field equations in beyond Einstein theories, which govern the dynamics of the compact objects and their perturbations, are also different from the Einstein equations. 
These differences entail that the dynamics and GW signatures of BBH coalescence in modified gravity  are different from those in GR. 
Thus, by comparing the measured GWs with the prediction of GR, we can put GR to the test in the strong field and either detector or constrain any deviations.  

As we pointed out in \cite{Chung:2024vaf}, GWs emitted during the ringdown stage have several advantages when testing GR. 
First, GWs during the ringdown stage usually have the highest frequencies as compared to other stages of BBH coalescence. 
These GW frequencies are related to the relative rates of change of gravitational perturbations; thus, their rapid change during the ringdown could reveal effects that might not be detectable in other stages. 
Second, ringdown waves are relatively easy to characterize because the QNM frequencies depend solely on the properties of the remnant BH and not its progenitors.
Third, the QNM response probes the geometry of the BH near its horizon, the region of spacetime where the field strength is greatest, and where deviations from GR may be most likely to manifest. These features make the ringdown stage an exceptional probe for testing the validity of GR in the strong field.

Most current ringdown tests, however, have been cast as null tests of GR,
which have advantages and disadvantages. While such tests are very general and agnostic, they lack the ability to connect constraints to actual physics or physical theories\footnote{This is unlike in the inspiral stage of BBH coalescence, where the parametrized post-Einsteinian framework~\cite{Loutrel:2014vja, Loutrel:2022xok} is both agnostic and able to connect constraints to specific physics.}. Tests that constrain specific beyond-Einstein models allow us to ``forward-model'' the ringdown and then measure or constrain the coupling constants of the specific model directly from the GW data. Moreover, by focusing on specific models we can learn what features in the ringdown GW response are generic and which are model specific, so that more sophisticated model-agnostic tests can be developed in the future. 

Two recently developed approaches are capable of computing QNM spectra in modified gravity, making model-specific ringdown tests of modifications to GR possible. 
The first is the modified Teukolsky formalism \cite{Li:2022pcy, Wagle:2023fwl, Li:2023ulk, Cano:2023tmv, Hussain:2022ins}, which derives and solves a modified Teukolsky equation for theories with leading-order deviations from GR. 
The second is the \textit{Metric pErTuRbations wIth speCtral methodS} (METRICS formalism), which two of us developed in \cite{Chung:2023wkd, Chung:2023zdq, Chung:2024ira, Chung:2024vaf}, with variants and extensions developed recently by others \cite{Blazquez-Salcedo:2023hwg, Blazquez-Salcedo:2024oek}. 
In METRICS, the linearized field equations are converted into homogenous algebraic equations whose eigenvalues and eigenvectors correspond to the QNM frequencies and perturbation profiles around the background BH, respectively. 
We applied METRICS to Einstein-scalar-Gauss-Bonnet (EsGB) gravity in \cite{Chung:2024ira, Chung:2024vaf}, a well-motivated extension of GR that arises as the low-energy limit of string theory \cite{Maeda:2009uy, Moura:2006pz}. 
This led to the first accurate computations of QNM spectra for rotating BHs with scalar hair and (dimensionless) spins of $a \sim 0.85$. 
The METRICS EsGB spectra were applied to an inspiral-merger-ringdown test of EsGB gravity in \cite{Julie:2024fwy}, leading to a stringent constraint on this effective theory.

In this work, we extend METRICS to dynamical Chern-Simons (dCS) gravity, another well-motivated modified gravity effective theory that can be interpreted as another sector of the low-energy/low-curvature limit of heterotic string theory~\cite{Bergshoeff:1981um, Green:1984sg} and loop quantum gravity, as an effective theory extension to GR, and is also motivated by gravitational anomalies in particle physics \cite{Alexander:2009tp}.
We begin by recapping the field equations, the background BH metric and the background scalar-field profile, and the mathematical form of the metric and scalar-field perturbations in dCS gravity (Sec.~\ref{sec:MGTs}). 
Using this information, we derive the linearized field equations, which we then transform into homogenous, algebraic equations through spectral expansions (Sec.~\ref{sec:Linear_field_eqns}). 
However, unlike in the case of EsGB gravity, we identify an issue that makes our perturbative method (developed in~\cite{Chung:2024ira, Chung:2024vaf}) to find a solution to the algebraic equations fail in dCS gravity (Sec.~\ref{sec:issues}).  
Instead, we compute the QNM frequencies at given values of the coupling constant of dCS gravity by solving the linearized field equations via Newton-Raphson iterations (as detailed in Sec.~\ref{sec:Newton_Raphson}).
Using this iteration scheme, we compute the QNM frequencies of the 022, 033, and 032 modes of rotating BHs in dCS gravity with dimensionless spin $ \in [0, 0.753]$ for a set of small dimensionless coupling parameters (Secs.~\ref{sec:Results-1} and~\ref{sec:Results-2}). 
Using these frequencies, we estimate the leading-order modifications to the GR QNM frequencies (Sec.~\ref{sec:Intermediate_spinning}) with a numerical uncertainty of $\lesssim 10^{-3}$ for all modes and computed dimensionless spin, and construct their optimal fitting polynomials (Sec.~\ref{sec:fitting_exp}). 
The leading-order frequency shifts in dCS gravity (Fig.~\ref{fig:omega_1_022} and Tables~\ref{tab:omega_1_022}, \ref{tab:omega_1_033} and \ref{tab:omega_1_021}) and their optimal fitting polynomial (Table~\ref{tab:poly_fit_coeffs}) are key results of this paper. 
Using the dCS frequencies obtained using METRICS, we project constraints on dCS gravity that could be obtained from future BBH GW data detected by the LIGO-Virgo-KAGRA observatories (Sec.~\ref{sec:DA}). 
We conclude this paper by examining the implications of this work on fundamental physics, and possible future work (Sec.~\ref{sec:Conclusions}). 

The QNM frequencies obtained in this work significantly enhance our understanding about the strong-field dynamics in dCS gravity. 
In several ways, we knew less about dCS gravity than EsGB gravity. 
Before this work, the QNM spectra have been computed for BHs of $a \leq 0.1$ in dCS gravity. 
Whereas there are several approaches to simulate compact binary coalescence in EsGB gravity \cite{Witek:2018dmd, East:2020hgw, Corman:2024vlk, Okounkova_Stein_Moxon_Scheel_Teukolsky_2020, Elley:2022ept, AresteSalo:2022hua, AresteSalo:2023hcp, Okounkova:2020rqw}, numerical simulations of compact objects in dCS gravity have only been performed using an order-reduction scheme \cite{Okounkova_Stein_Scheel_Hemberger_2017, Okounkova:2019dfo}. 
Due to numerical errors, it is difficult to accurately extract QNM frequencies from these simulations. 
The frequencies computed in this work fill this gap exactly, offering us additional insight into the spacetime dynamics in dCS gravity theory. 

The work described in this paper is largely based on the spectral code METRICS, whose details and numerical implementations are described in previous publications \cite{Chung:2023zdq, Chung:2023wkd, Chung:2024ira, Chung:2024vaf}. 
In this paper, to avoid repetition, only the necessary results relevant to this work will be recapped in a summarized form, and the details of these results can be traced back to the previous METRICS publications. 
Reading the previous METRICS publication will greatly facilitate the reader's comprehension of this paper. 
Henceforth, we assume the following conventions: 
$x^{\mu} = (x^0, x^1, x^2, x^3) = (t, r, \chi, \phi)$, where $\chi = \cos \theta$ and $\theta$ is the polar angle;
the signature of the metric tensor is $(-, +, +, +)$;
gravitational QNMs are labeled by $n l m$ or $(n, l, m)$, where $n$ is the principal mode number, $l$ is the azimuthal mode number 
and $m$ is the magnetic mode number of the QNM;
the QNM frequencies computed using the METRICS approach are referred to as ``METRICS QNM frequencies"; 
Greek letters in index lists stand for spacetime coordinates.

\section{Black-hole perturbations in Dynamical Chern-Simons Gravity}
\label{sec:MGTs}

In this section, we review dCS gravity. 
We begin by describing the field equations and their solution for stationary and axisymmetric BH spacetimes. 
We then discuss how we perturb these backgrounds.  
\subsection{The field equations}
\label{sec:FEs}

The Lagrangian density of dCS gravity can be written as \cite{Yagi:2015oca, Nair:2019iur, Perkins:2021mhb}
\begin{equation}\label{eq:Lagrangian}
16 \pi \mathscr{L} = R - \frac{1}{2} \nabla_{\mu} \Phi \nabla^{\mu} \Phi- V(\Phi) + \alpha f(\Phi) \mathscr{P}, 
\end{equation}
where $\Phi$ is a scalar field to which the BH in the modified gravity couples, $V(\Phi)$ is the potential of $\Phi$, $\alpha$ is a coupling constant, which characterizes the strength of the modifications to gravity and has dimensions of length squared in geometric units \footnote{The coupling constant in this paper follows the conventions in \cite{Cano_Ruiperez_2019}, and it is four times the coupling constant in \cite{Wagle:2021tam, Wagle:2023fwl, isospectrality-paper, QNM_dCS_02}.}, and $f(\Phi)$ is a function of $\Phi$ only, and $\mathscr{P}$ is the Pontryagin density, defined as
\begin{equation}
\mathscr{P} = R_{\mu \nu \rho \sigma} \tilde{R}^{\mu \nu \rho \sigma}, 
\end{equation}
where $\tilde{R}^{\mu \nu \rho \sigma}$ is the dual Riemann tensor
\begin{equation}
\tilde{R}^{\mu \nu \rho \sigma} = \frac{1}{2} \varepsilon^{\rho \sigma \alpha \beta} R^{\mu \nu} {}_{\alpha \beta}, 
\end{equation}
$\varepsilon^{\rho \sigma \alpha \beta}$ is the Levi-Civita tensor, defined as 
\begin{equation}
\varepsilon^{\rho \sigma \alpha \beta} = \frac{1}{\sqrt{-g}} [\rho \sigma \alpha \beta], 
\end{equation}
$g$ is the determinant of $g_{\mu \nu}$, and $[\rho \sigma \alpha \beta]$ is the totally asymmetric Levi-Civita symbol. 
In this paper, in accordance with \cite{Cano_Ruiperez_2019}, we adopt the following convention of the Levi-Civita symbol\footnote{Different literature might adopt a different convention of the Levi-Civita symbol. 
We advise the reader to thoroughly check the convention before making reference to the equations in the literature. }
\begin{equation}
[0123] = [t r \chi \phi] = + 1. 
\end{equation}

Because of the motivation explained in \cite{Chung:2024ira, Chung:2024vaf}, in this work, we focus on the cases of zero potential and a shift-symmetric coupling function, 
\begin{equation}\label{eq:shift-symmetric}
\begin{split}
V(\Phi) &= 0, \qquad
f(\Phi) = \Phi. 
\end{split}
\end{equation}
As pointed out in \cite{Chung:2024vaf}, a shift-symmetric coupling function can be viewed as the small-coupling approximation (or limit, i.e. when $\alpha \ll 1 $) of a general coupling function because $\mathscr{P}$ is a topological invariant.  

Given the Lagrangian density, one can use the least action principle to derive the field equations of dCS gravity in vacuum, which can be schematically expressed as 
\begin{align}
\label{eq:field_eqs}
& R_{\mu}{}^{\nu} + \zeta \left( \mathscr{A}_{\mu}{}^{\nu} - T_{\mu}{}^{\nu} \right) = 0, \\
\label{eq:field_eqs_scalarfield}
& \square \vartheta + \mathscr{P} = 0, 
\end{align}
where $\zeta$ is a dimensionless coupling parameter, 
\begin{equation}\label{eq:zeta}
\zeta = \frac{\alpha^2}{M^4}, 
\end{equation}
with $M$ the BH mass, 
$\mathscr{A}_{\mu}{}^{\nu} = g_{\mu \alpha} \mathscr{A}^{\alpha \nu}$, and $\mathscr{A}^{\mu \nu}$ defined by \cite{dCS_01, dCS_02, dCS_03, QNM_dCS_01, QNM_dCS_02, QNM_dCS_03, QNM_dCS_04}, 
\begin{align}\label{eq:Amunu_dCS}
- \frac{1}{4}\mathscr{A}^{\mu \nu} & \equiv \left(\nabla_\sigma \vartheta \right) \varepsilon^{\sigma \delta \alpha(\mu|} \nabla_\alpha R^{|\nu)}{}_{\delta} + \left(\nabla_\sigma \nabla_\delta \vartheta\right) \tilde{R}^{\delta (\mu \nu) \sigma}. 
\end{align}
The quantity $\vartheta$ is a \textit{rescaled} scalar field, such that $\Phi = \alpha \vartheta$, and 
\begin{equation}\label{eq:Tmunu}
\begin{split}
T_{\mu}{}^{\nu} & \equiv \frac{1}{2}\left(\nabla_{\mu} \vartheta \right)\left(\nabla^{\nu} \vartheta \right), \\
\end{split}
\end{equation}
is the trace-reversed energy-momentum tensor of the rescaled scalar field. 

\subsection{Background spacetime and scalar field}
\label{sec:bkgd}

To study perturbations of rotating BHs in dCS gravity, we must first construct a stationary, axisymmetric, and vacuum rotating BH spacetime, which requires solving the field equations [Eq.~\eqref{eq:field_eqs}].
Since Eq.~\eqref{eq:field_eqs} is a complicated set of nonlinear partial differential equations, we seek to solve for the background metric and scalar field as a power series in $a$, following the approach in~\cite{Yunes:2009hc,Yagi:2012ya,Cano_Ruiperez_2019}. 
Specifically, the background scalar field takes the following form
\begin{equation}\label{eq:vartheta_fns}
\vartheta (r, \chi) = \sum_{k=0}^{} \sum_{p=0}^{N_{r}} \sum_{q=0}^{N_{\chi}} \vartheta_{i, k, p, q} \frac{a^{k} \chi^{q} }{r^p}, 
\end{equation}
where $\chi = \cos{\theta}$ and ${\vartheta}_{i, k, p, q}$ are constant, and $N_{r}(k)$ and $N_{\chi}(k)$ are a positive integer that depend on $k$. 
We use Boyer-Lindquist coordinates to construct the solutions because, in these coordinates, the radial position of the event horizons remains the same as in GR. 
In Boyer-Lindquist coordinates, the dCS background metric takes the following form \cite{Cano_Ruiperez_2019, Cano:2023jbk}, 
\begin{widetext}
\begin{equation}\label{eq:metric}
\begin{split}
ds^2 &= g_{\mu \nu}^{(0)} dx^{\mu}  dx^{\nu} = - \left( 1-\frac{2 M r}{\Sigma} - \zeta H_1(r, \chi)\right) dt^2 - \left[ 1 + \zeta H_2(r, \chi) \right] \frac{4 M^2 a r}{\Sigma} (1 - \chi^2) d \phi dt \\
& \quad + \left[ 1 + \zeta H_3(r, \chi) \right] \left( \frac{\Sigma}{\Delta} dr^2 + \frac{\Sigma}{1 - \chi^2} d \chi^2 \right) + \left[ 1 + \zeta H_4(r, \chi) \right](1-\chi^2) \left[r^{2} + M^2 a^{2}+\frac{2 M^3 a^{2} r}{\Sigma} (1 - \chi^2)\right] d\phi^2, \\
\end{split}
\end{equation}
\end{widetext}

where $\Sigma = r^2 + M^2 a^2 \chi^2$, $\Delta = (r-r_+)(r-r_-)$, and $r_{\pm} = M(1 \pm \sqrt{1-a^2})$. 
The quantities $H_{i=1, 2, 3, 4}$ are dCS corrections to the Kerr metric, which can be expressed as a power series in $a$, 
\begin{equation}\label{eq:H_fns}
H_i (r, \chi) = \sum_{k=0} \sum_{p=0}^{N_{r}'(k)} \sum_{q=0}^{N_{\chi}'(k)} h_{i, k, p, q} \frac{a^{k} \chi^{q} }{r^p}, 
\end{equation}
where $h_{i, k, p, q}$ are constants, and $N_{r}'(k)$ and $N_{\chi}'(k)$ are positive integers that depend on $k$. 
In dCS gravity, the angular velocity and surface gravity of the event horizon of rotating BHs  are modified by an amount $\sim \mathcal{O}(\zeta)$. 
The explicit power series expression for the leading-in-$\alpha$ corrections to the angular velocity, surface gravity, $\vartheta$ and $H_{i=1,...,4}$ as a power series up to the 40th order of $a$ are given in a Mathematica notebook that is available upon request. 

We are here interested in calculating the QNM frequencies of perturbations of rotating BHs with large spins, but we will employ the power-series expansion in $a$ to represent the background. This procedure is valid provided we keep sufficiently large $N_{r}(k)$, $N_{\chi}(k)$, $N_{r}'(k)$ and $N_{\chi}'(k)$, so that the approximate (series-expanded-in-$a$) background solution is sufficiently close to the exact solution, which is only known numerically. We have here ensured that the error we introduce by truncating these approximate series solution is much smaller than other numerical errors in the calculation of the QNM frequencies. 

\subsection{Perturbations of fields}
\label{sec:metpert}

We now consider both metric and scalar perturbations of a BH in dCS gravity, 
\begin{equation}\label{eq:metpert}
\begin{split}
g_{\mu\nu} & = g_{\mu\nu}^{(0)} + \varepsilon \; h_{\mu\nu} \,,\\
\vartheta(r, \chi) & = \vartheta^{(0)}(r, \chi)  + \varepsilon e^{im\phi-i\omega t} h_7 (r, \chi), 
\end{split}
\end{equation}
where $g_{\mu\nu}^{(0)}$ is the background metric of the rotating BHs, $\vartheta^{(0)}(r, \chi)$ is the background, rescaled scalar field, $h_{\mu\nu}$ and $h_7$ are respectively the metric and scalar-field perturbations, and $\varepsilon$ is a bookkeeping parameter for the perturbations. 
To simplify our calculations, we enforce the Regge-Wheeler gauge, which we have checked can be enforced in this gravity theory \cite{Wagle:2019mdq}. 
In this gauge, $h_{\mu\nu}$ can be written as~\cite{Regge:PhysRev.108.1063, Berti_02}
\begin{widetext}
\begin{equation}\label{eq:metpert}
h_{\mu\nu} (t,r,\chi,\phi) = e^{im\phi-i\omega t}
\begin{pmatrix}
    h_1(r,\chi) & h_2(r,\chi) & -im(1-\chi^2)^{-1} h_5(r,\chi) &  (1-\chi^2) \partial_\chi h_5(r,\chi) \\
    * & h_3(r,\chi) & -im(1-\chi^2)^{-1} h_6(r,\chi) &  (1-\chi^2) \partial_\chi h_6(r,\chi) \\
	* & * & \left( 1 - \chi^2 \right)^{-1} h_4(r,\chi) & 0  \\
	* & * & * & \left( 1 - \chi^2 \right) h_4(r,\chi)  
\end{pmatrix}, 
\end{equation}
\end{widetext}
where the asterisks represent symmetrization. Henceforth, it is to be  understood that $\omega$ depends on $(n,l,m)$, although we suppress its indices here for simplicity.

During the ringdown phase, GWs are purely ingoing at the event horizon and purely outgoing at spatial infinity. 
These boundary conditions imply that GW amplitude diverges in these two limits. 
Mathematically, to accommodate this asymptotically divergent behavior, we write $h_k(r, \chi)$ as
\be \label{eq:radspec}
h_k(r, \chi) = A_k(r)  u_k(r, \chi) \,. 
\ee
Here $A_k(r)$ is an asymptotic factor, which we have determined in \cite{Chung:2024vaf} to be, 
\begin{equation}\label{eq:asym_prefactor}
\begin{split}
A_k(r) = & e^{i \left(1 + \frac{1}{2} \zeta H_3^{(0)} \right)\omega r} r^{2 i M \omega  + \rho_{\infty}^{(k)}} \left( \frac{r-r_+}{r}\right)^{-i\frac{\omega - m \Omega_{\rm H}}{2 \kappa}- \rho_H^{(k)}}\,,
\end{split}
\end{equation}
$\rho_{H}^{(k)}$ and $\rho_{\infty}^{(k)}$ are respectively a $k$-dependent index that control the divergence of $h_{k}$ near the event horizon and spatial infinity, and $H_3^{(0)} = \lim_{r \rightarrow + \infty} H_3(r)$. We have check that this asymptotic factor continues to be the correct choice for dCS gravity. 
In \cite{Chung:2024vaf}, we have determined $\rho_{H}^{(k)}$ to be 
\begin{equation}\label{eq:rhos}
\begin{split}
\rho_{H}^{(k)} & = 
\begin{cases}
2, ~~ \text{for $k \neq 4$ nor 7,}\\
1, ~~ \text{for $k = 4$,} \\
-1, ~~ \text{for $k = 7$,} \\
\end{cases}
\end{split} 
\end{equation}
and $\rho_{\infty}^{(k)}$ to be 
\begin{equation}\label{eq:rhos}
\begin{split}
\rho_{\infty}^{(k)} & = 
\begin{cases}
2, ~~ \text{for $k \neq 4$ nor 7,}\\
1, ~~ \text{for $k = 4$,} \\
-1, ~~ \text{for $k = 7$.} \\
\end{cases}
\end{split} 
\end{equation}
The function $u_k(r,\chi)$ is a correction factor that is finite, bounded and unknown; this is the function we will spectrally expand in the next section and then solve for using the modified field equations. 

\section{Linearized field equations}
\label{sec:Linear_field_eqns}

The linearized field equations in dCS gravity are the central equations that we solve for the QNM frequencies with METRICS. 
In this section, we derive the linearized field equations that govern the field perturbations ($h_k$) and the finite correction functions ($u_k$). 
Then, we explain the detailed procedure of applying METRICS to these linearized field equations. 

\subsection{Simplifications of the linearized field equations}
\label{sec:simplication_linear_field_eqns}

Substituting the perturbed fields (Eq.~\eqref{eq:metpert}) into the field equations in dCS gravity, and expanding the perturbed field equations up to first order in $\varepsilon$ and $\zeta$, we obtain the linearized field equations. 
Naturally, the field equations contain many terms. 
To simplify the linearized field equations, we invoke the small-coupling approximation. 
We recall that the Lagrangian density in Eq.~\eqref{eq:Lagrangian}, the field equations [Eq.~\eqref{eq:field_eqs}], and the background metric in Eq.~\eqref{eq:metric} are valid only up to the first order in $\zeta$. 
Thus, in this work, we focus on computing QNM frequencies that are also accurate only to first order in $\zeta$. We stress that it is \textit{mathematically and physically inconsistent to calculate the QNM frequencies to all orders in $\zeta$ when the theory under study is an effective field theory} (like dCS gravity), which is naturally a truncated derivative expansion at the level of the action. If one wishes to understand the behavior of QNM frequencies (or any other observable) to higher order in $\zeta$, then higher order terms in the action must be included. 

We can use the effective-field-theory nature of dCS gravity to simplify some of the resulting linearized field equations. 
As far as only the first-$\zeta$ order shift of the QNM frequencies is concerned, $\mathscr{A}_{\mu} {}^{\nu}$, $T_{\mu} {}^{\nu}$, the scalar field equation, and their perturbations can be computed in the GR Kerr background. 
Thus, in the small-coupling limit, the term 
\begin{equation}\label{eq:simplification}
\left(\nabla_\sigma \vartheta \right) \varepsilon^{\sigma \delta \alpha(\mu|} \nabla_\alpha R^{|\nu)}{}_{\delta}
\end{equation}
in $\mathscr{A}^{\mu \nu}$ and its perturbations vanish \cite{Wagle:2023fwl}. 
In other words, effectively, 
\begin{align}\label{eq:Amunu_simplified}
\mathscr{A}^{\mu \nu} & = - 4 \left(\nabla_\sigma \nabla_\delta \vartheta\right) \tilde{R}^{\delta (\mu \nu) \sigma}. 
\end{align}
This simplification can significantly reduce the length of the linearized field equations and the time needed for computations of the QNM frequencies. 

\subsection{Conversion of the linearized field equations into homogenous algebraic equations via spectral expansions}
\label{sec:algebraic_eqns}

The linearized field equations in dCS gravity theory are a set of linear partial differential equations that, to linear order in $\zeta$, are second order in $h_{i=1, ..., 4, 7}$ but third order\footnote{We stress that this is not because the field equations are third order, since the third order term can be discarded in the effective field theory expansion. Rather, the third derivatives here appear because the metric perturbation is defined in terms of first derivatives of $h_{i=5,6}$, as shown in Eqs.~\eqref{eq:metpert}.} in $h_{i=5,6}$ 
To solve these equations via METRICS, we define a compactified radial coordinate \cite{Langlois:2021xzq, Jansen:2017oag}
\begin{equation}\label{eq:z}
z(r) = \frac{2r_+}{r} - 1,   
\end{equation}
such that $z \in (-1, +1)$. 
In this coordinate, we can perform a spectral expansion on $u_k$ as 
\begin{equation}
\label{eq:spectral_decoposition_correction_factor}
u_k (z, \chi) = \sum_{n=0}^{\infty} \sum_{\ell=|m|}^{\infty} v_k^{n \ell}  T_{n}(z) P^{|m|}_{\ell}(\chi)\,. 
\end{equation}
Here $T_n(z)$ are Chebyshev polynomials of $n$-th order and $P^{|m|}_{\ell}$ is the associated Legendre polynomials of order $|m|$ and degree $\ell$. 

Observe that the background metric, scalar field, and $z(r)$ involve only rational functions of $r$ and $\chi$. Thus, the linearized field equations can be brought into a form whose coefficients functions are a polynomial of $z$ and $\chi$, via the procedures described in \cite{Chung:2023zdq, Chung:2023wkd, Chung:2024vaf}, namely
\begin{equation}\label{eq:system_3}
\begin{split}
& \sum_{j=1}^{6} \sum_{\alpha, \beta = 0}^{\alpha + \beta \leq 3} \sum_{\gamma=0}^{2} \sum_{\delta=0}^{d_{z}} \sum_{\sigma=0}^{d_{\chi}} \mathcal{K}_{k, \gamma, \delta, \sigma, \alpha, \beta, j} \omega^\gamma z^{\delta} \chi^{\sigma} \partial_{z}^{\alpha} \partial_{\chi}^{\beta} u_j = 0 \,. 
\end{split}
\end{equation}
Here $\mathcal{K}_{k, \alpha, \beta, \gamma, \delta, \sigma, j}$ are constants, $d_z$ and $d_{\chi}$ are the degree of $z$ and $\chi$ of the coefficient of the partial derivative $\partial_{z}^{\alpha} \partial_{\chi}^{\beta} \{...\} $ in the equations respectively. 
Substituting Eq.~\eqref{eq:spectral_decoposition_correction_factor} into Eq.~\eqref{eq:system_3}, and then performing spectral expansions to both sides of the equations, we arrive at a set of linear homogeneous algebraic equations of $v_k^{n \ell}$. 
Let us denote a vector which contains all $v_k^{n \ell}$ by $\textbf{v}$, so that the resulting algebraic equations can be written as \cite{Chung:2023zdq, Chung:2023wkd, Chung:2024ira, Chung:2024vaf}
\begin{equation}\label{eq:augmented_matrix_01}
\begin{split}
\tilde{\mathbb{D}} (\omega) \textbf{v} = \left[ \tilde{\mathbb{D}}_0 + \tilde{\mathbb{D}}_1 \omega + \tilde{\mathbb{D}}_2 \omega^2 \right] \textbf{v} = \textbf{0} \,,
\end{split}
\end{equation}
where the $\tilde{\mathbb{D}}_{0,1,2}$ matrices are constant, $ 11 (\mathcal{N}_z + 1)(\mathcal{N}_{\chi} + 1) \times 7 (\mathcal{N}_z + 1)(\mathcal{N}_{\chi} + 1) $ rectangular matrices.
The QNM frequencies of a rotating dCS BH correspond to the $\omega$ such that Eq.~\eqref{eq:augmented_matrix_01} admits a nontrivial solution $\textbf{v}$.

\subsection{Issues with perturbative solutions to the spectral eigenvalue equations in dCS gravity}
\label{sec:issues}

When attempting to solve the above spectral eigenvalue, algebraic equations using perturbation theory, however, we encountered certain problems. We developed this perturbative solution method in \cite{Chung:2024ira, Chung:2024vaf} and applied it successfully in EsGB gravity, but the same implementation fails in dCS gravity. 
Specifically, we observed that the backward modulus difference of the leading-order frequency shift failed to decrease steadily with increasing spectral order, even for a slowly rotating BH. Such a steady decrease in the backward modulus difference with spectral order is a necessary condition for the stability of the perturbative solution method. 

We have identified the following potential cause for this issue. 
Since the Pontryagin density involves the totally asymmetric Levi-Civita symbol, its linearization in terms of $h_{i=1, ..., 6}$ contains much fewer terms than the linearization of the Gauss-Bonnet invariant. 
This feature results in more zero elements in the rows of $\tilde{\mathbb{D}} (\omega)$ that the corresponding linearization of $ \Box \vartheta + \mathscr{P} = 0 $ at $\zeta = 0$. 
The larger number of zero elements may make computing the Moore-Penrose inverse, which is essential for the spectral-eigenvalue perturbations (see \cite{Chung:2024ira, Chung:2024vaf}), unstable. 

\subsection{Newton-Raphson iterations}
\label{sec:Newton_Raphson}

Instead of using the perturbative solution scheme of~\cite{Chung:2024ira, Chung:2024vaf}), we here compute the QNM frequencies for a given small value of $\zeta$ through the use of a Newton-Raphson method. 
The linearized field equations (Eq.~\eqref{eq:system_3}) are a set of homogenous partial differential equations. 
At a given QNM frequency $\omega$, the equations can admit many solutions, each proportional to the others \cite{Chung:2023zdq}. 
To ensure our numerical computation can converge to a particular solution, we adopt the following convention for perturbations of different sectors, following the previous METRICS computations in~\cite{Chung:2023wkd, Chung:2024ira, Chung:2024vaf}. 
For axial perturbations, we demand
\begin{equation}\label{eq:polar_convention}
v_{k=1}^{n=0,\ell=|m|} = 1\,,
\end{equation}
while for polar perturbations we require that
\begin{equation}\label{eq:polar_convention}
v_{k=5}^{n=0,\ell=|m|} = 1\,,
\end{equation}
and for scalar perturbations we ask that 
\begin{equation}\label{eq:polar_convention}
v_{k=7}^{n=0,\ell=|m|} = 1\,. 
\end{equation}
These conventions can always be enforced because of the homogeneity of Eq.~\eqref{eq:system_3} \cite{Chung:2024vaf}. 
For any solution whose corresponding component is not one, we can always divide all $v_{k}^{n,\ell} $ by the corresponding component to make the solution consistent with this convention. 

By enforcing this convention, we have removed one unknown. 
The remaining unknowns are the remaining components of $v_{k}^{n, \ell}$ and $\omega$. 
Let us denote these unknowns using the vector $\textbf{x}$, and write it as 
\begin{equation}
\textbf{x} = \left\{ \hat{\textbf{v}}, \omega \right\}, 
\end{equation}
where $\hat{\textbf{v}}$ denotes the remaining components of $\textbf{v}$, and $\omega$ is the QNM frequency of the corresponding sector. 
To initialize the Newton-Raphson iterations, we also need an initial guess. 
Following previous METRICS computations, we pick $\hat{\textbf{v}} = 0 $ and $\omega = \omega^{(0)}$, the unmodified QNM frequency in GR, as our initial guess \cite{Chung:2024vaf}. 
The initial guess for $\omega$ is justified because we have shown that METRICS can converge to the correct GR frequency with a displaced initial guess in \cite{Chung:2024vaf}. 
Also, since $\zeta \ll 1$, we expect that $\omega - \omega^{(0)} \sim \mathcal{O}(\zeta)$. 
This initial choice of $\omega$ therefore speeds up our computations. 

Let us now denote the left-hand side of the first equal sign in Eq.~\eqref{eq:augmented_matrix_01} evaluated at a given $\textbf{x}$, with the corresponding component having been set to 1, by $\textbf{f}(\textbf{x})$. 
The goal of the Newton-Raphson iterations is to numerically solve $\textbf{f}(\textbf{x}) = \textbf{0}$ for $\textbf{x}$. 
To this end, we update our guess via
\begin{equation}
\textbf{x}_{n+1} = \textbf{x}_{n} - \textbf{J}^{-1} \cdot \textbf{f}(\textbf{x}_{n}), 
\end{equation}
where $\textbf{x}_{n+1}$ and $\textbf{x}_{n}$ are respectively the guess in the $(n+1)$th and $n$th iterations, $\textbf{J}$ is the $11 (\mathcal{N}_{z}+1)(\mathcal{N}_{\chi}+1) \times 7 (\mathcal{N}_{z}+1)(\mathcal{N}_{\chi}+1)$ Jacobian matrix, whose $(i,j)$th element is given by 
\begin{equation}
[\textbf{J}]_{ij} = \frac{\partial f_i}{\partial [\textbf{x}]_{j}} \Bigg|_{\textbf{x} = \textbf{x}_{(n)}}, 
\end{equation}
and the superscript $-1$ denotes the Moore-Penrose inverse.  

The Newton-Raphson method does not solve $ \textbf{f}(\textbf{x}) = \textbf{0}$ exactly. 
Rather, the iterations at best converge to an approximate numerical solution that satisfies this vector equation to a given error tolerance. 
We terminate the iterations when the error tolerance is achieved. 
In this work, we terminate the iterations when
\begin{equation}\label{eq:tolerance_error}
\| \textbf{f}(\textbf{x}_{n}) \|_{2} < \epsilon, 
\end{equation}
where $ \| \textbf{f}(\textbf{x}_{n}) \|_{2} $ is the $L^2$ norm of the residual vector $\textbf{f}(\textbf{x}_{n})$ and $\epsilon$ is the error tolerance. 
As in previous METRICS calculations \cite{Chung:2023wkd, Chung:2024ira, Chung:2024vaf}, we set $\epsilon = 10^{-7}$ in this work, and the inverse of the Jacobian matrix is computed using the built-in \texttt{PseudoInverse} function of \textit{Mathematica} to double precision. 

\section{Numerical Results for slowly rotating black holes}
\label{sec:Results-1}

We here present the results obtained by implementing the above Newton-Raphson method to solve the linearized, algebraic equations for slowly rotating black holes. 
We first describe results when the background is not spinning, and then move on to very slowly spinning black holes. 
This study allows us to gain insight into the structure of QNMs in dCS and to check our results against previous work. 

\subsection{QNM frequencies of nonrotating ($a = 0$) black holes}

\begin{figure*}[htp!]
\centering  
\subfloat{\includegraphics[width=0.47\linewidth]{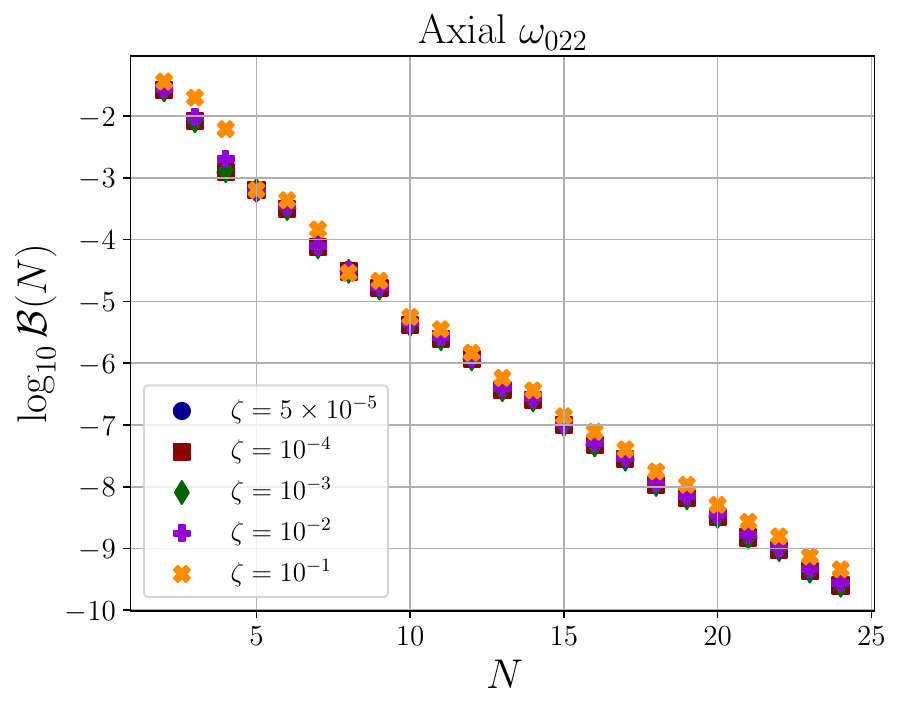}}
\subfloat{\includegraphics[width=0.47\linewidth]{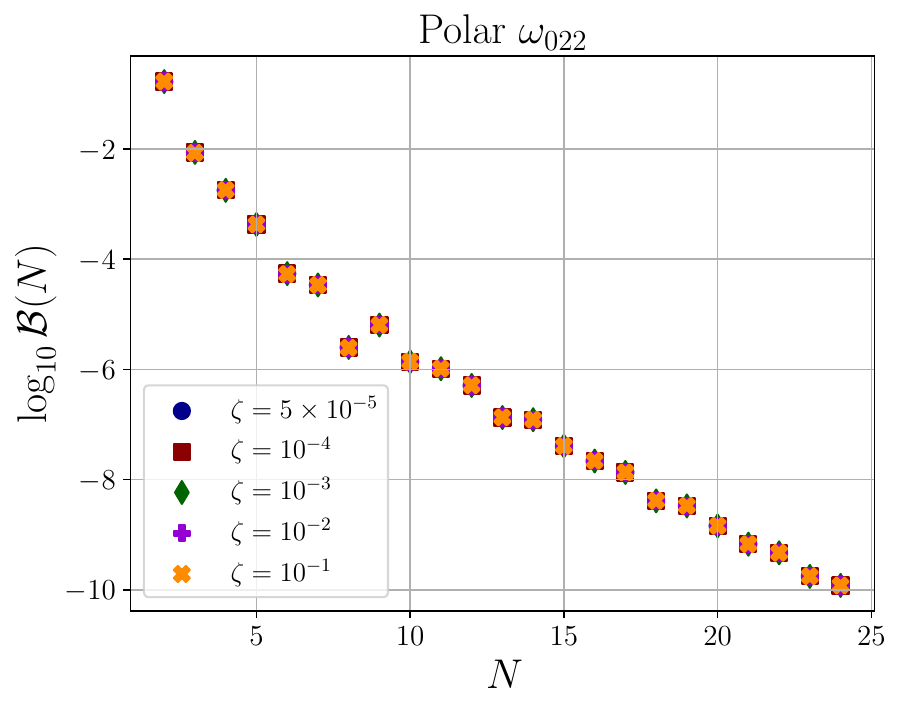}}
\caption{The backward modulus difference $\mathcal{B}(N) = |\omega^{(1)}(N)-\omega^{(1)}(N-1)|$, where $\omega^{(1)}$ is the $nlm=$022-mode frequency of the axial (left) and polar (right) perturbations of a nonrotating BH ($a=0$) in shift-symmetric dynamical Chern-Simons (dCS) gravity as a function of the spectral order ($N$) for different value of $\zeta$. 
We observe that for all $\zeta$, $\mathcal{B}(N)$ decreases exponentially, indicating exponential convergence of the METRICS frequencies. 
This exponential convergence is consistent with the fact that the Schwarzschild metric is an exact solution to the field equations in dCS gravity without small-spin expansion and small-coupling approximation. 
}
\label{fig:BWD_a_0}
\end{figure*}

We first validate our METRICS implementation in dCS gravity in the nonrotating case. 
When $a=0$, the Schwarzschild BH with $\vartheta \equiv 0 $ is an exact solution to the field equations.
Because of this exact nature of the solution, we expect that METRICS can converge to the QNM frequencies exponentially. 
Figure~\ref{fig:BWD_a_0} shows the backward modulus difference, defined henceforth as 
\begin{equation}\label{eq:BWD}
\mathcal{B} (N) = |\omega^{(1)}(N) - \omega^{(1)}(N-1)|, 
\end{equation}
of the 022-mode, axial and polar metric perturbation to the Schwarzschild BH in dCS gravity as a function of the spectral order $N$ for different values of $\zeta$. 
Observe that $\mathcal{B} (N)$ approaches zero exponentially for all $ \zeta$, confirming the expectation that METRICS converges exponentially. 
We also observe that $\mathcal{B} (N)$ of the polar 022 mode does not vary with $\zeta$, while $\mathcal{B} (N)$ of the axial 022 mode does. 
This is because the polar 022-mode frequency of the Schwarzschild BH is not modified by the dCS coupling, as shown initially in~\cite{Cardoso:2009pk, Molina:2010fb}. 

\begin{figure}[tp!]
\centering  
\subfloat{\includegraphics[width=\columnwidth]{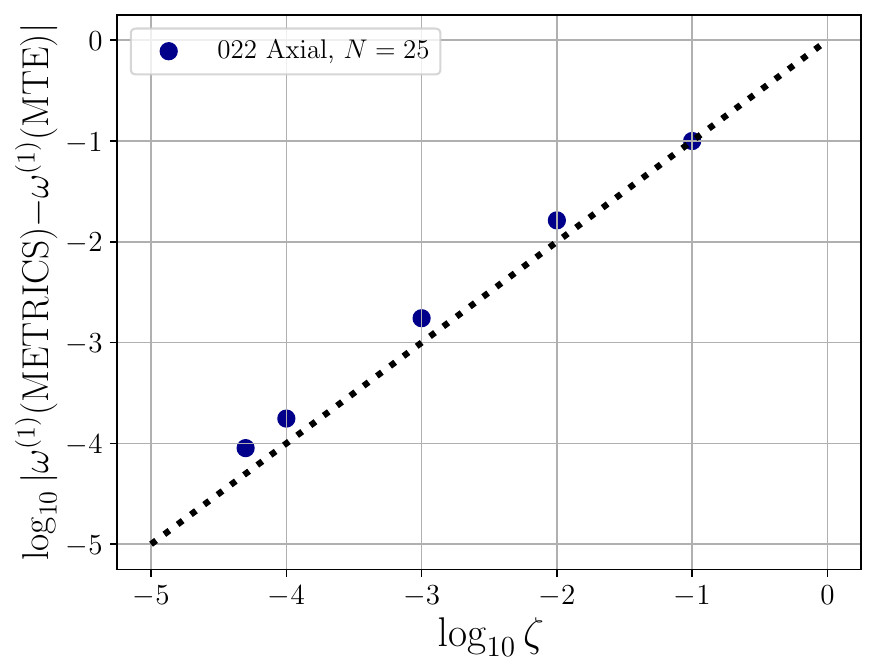}}
\caption{The modulus of the difference between the $\omega^{(1)}$, the leading-order QNM frequency shift of the axial perturbations to a nonrotating BH in dCS gravity, computed using METRICS using 25 spectral bases via Eq.~\eqref{eq:FWD} and that computed using the modified Teukolsky formalism, as a function of $\zeta$. 
We observe that the modulus difference is approximately proportional to $\zeta$, which is consistent with the error expected from the eigenvalue-perturbation scheme used by the modified Teukolsky formalism.}
 \label{fig:Diff_w1_MET_vs_MTE}
\end{figure}

Next, we estimate the error of our METRICS implementation in dCS gravity by comparing the METRICS 022-mode frequencies for the Schwarzschild BH in dCS gravity against previously known results.
The QNM frequencies of the polar perturbations to a Schwarzschild BH in dCS gravity are not modified, as mentioned above. 
Indeed, we observe that the METRICS frequencies of the 022-mode polar perturbations when $a=0$ converge to the GR frequencies progressively with $N$, as shown in the right panel of Fig.~\ref{fig:BWD_a_0}. 

As for the axial 022-mode frequency, we compare the METRICS results against the leading-order-in-$\zeta$ QNM frequency shift, $\omega^{(1)}$, obtained by solving the modified Teukolsky equation via an eigenvalue perturbation method \cite{Li:2025fci}. 
More precisely, we expand the frequencies as 
\begin{equation}\label{eq:w_zeta}
\omega(\zeta) = \omega^{(0)} + \omega^{(1)} \zeta, 
\end{equation}
where $\omega$ is the modified QNM frequency, $\omega^{(0)}$ is the GR frequency, and $\omega^{(1)}$ is the leading-order-in-$\zeta$ QNM frequency shift. The eigenvalue method applied to the modified Teukolsky formalism automatically yields the shift, which we denote $\omega^{(1)} (\rm MTE)$. The Newton-Raphson approach applied to the METRICS formalism does not yield the shift, but rather the full frequency in Eq.~\eqref{eq:w_zeta} plus higher-order terms in $\zeta$. We therefore calculate the METRICS leading-$\zeta$-order shift, $\omega(\zeta|\text{METRICS})$, via 
\begin{equation}\label{eq:FWD}
\omega^{(1)}(\text{METRICS}) = \frac{\omega(\zeta|\text{METRICS}) - \omega^{(0)}}{\zeta}, 
\end{equation}
where $\omega(\zeta|\text{METRICS})$ is the METRICS frequency computed at a given value of $\zeta$; specifically, we choose $\zeta = 5\times 10^{-5}, 10^{-4}, 10^{-3}, 10^{-2}$ and $10^{-1}$, and $N=25$, where $\mathcal{B}$ is minimized to compute $\omega(\zeta|\text{METRICS})$.

With this in hand, we now  compare $\omega^{(1)} (\rm METRICS)$ with $\omega^{(1)} (\rm MTE)$.
Figure~\ref{fig:Diff_w1_MET_vs_MTE} shows the base-10 log of $|\omega^{(1)} (\rm METRICS)-\omega^{(1)} (\rm MTE)|$ as a function of the base-10 log of $\zeta$. 
Observe that the difference between $\omega^{(1)} (\rm METRICS)$ and $\omega^{(1)} (\rm MTE)$ increases with $\zeta$ approximately linearly. 
This is because $\omega^{(1)} (\rm MTE)$ was computed only to linear order in $\zeta$, while  $\omega^{(1)} (\zeta|\rm METRICS)$ contains higher order in $\zeta$ corrections that should not have been retained. 
The difference is then of ${\cal{O}}(\zeta)$ because we have pulled out a factor of $\zeta$ in Eq.~\eqref{eq:w_zeta} for the linear shift. 
Figure~\ref{fig:Diff_w1_MET_vs_MTE} also shows that the 022-mode axial frequency of a Schwarzschild BH computed using METRICS is closely consistent with that computed using the modified Teukolsky formalism, apart from differences of ${\cal{O}}(\zeta^2)$. 
This consistency provides our first validation of the METRICS approach in dCS gravity. 

\begin{figure}[tp!]
\centering  
\subfloat{\includegraphics[width=\columnwidth]{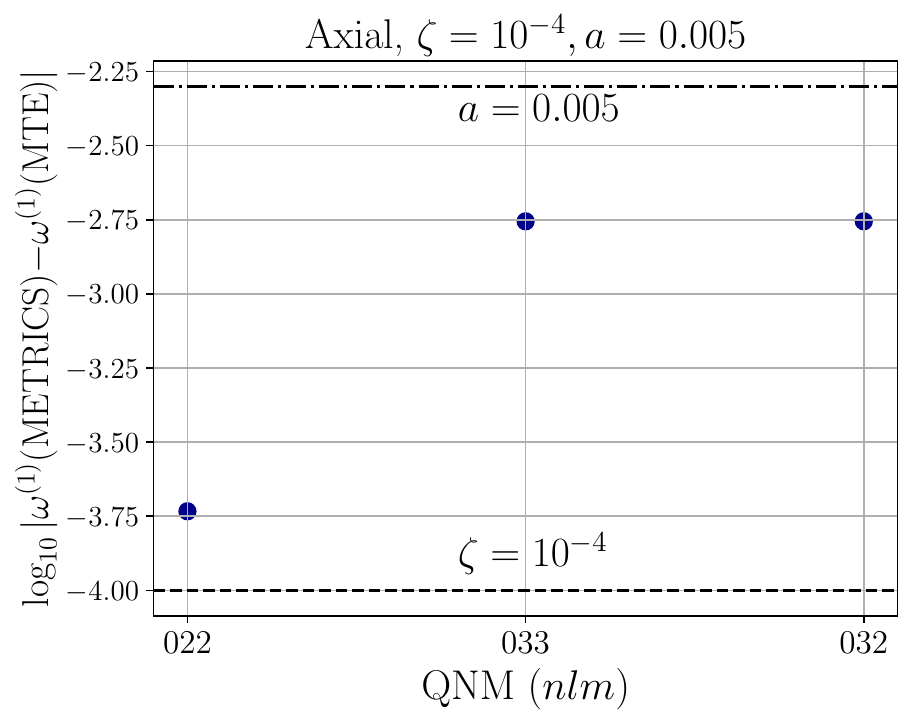}}
\caption{The modulus difference between the leading-$\zeta$-order shift ($\omega^{(1)}$) of the QNM frequency of the axial metric perturbations of the 022, 032, and 033 modes of a rotating BH of $a=0.005$ computed using METRICS [$\omega^{(1)}(\rm METRICS)$] and that using the modified Teukolsky equation ($\omega^{(1)}$), assuming $\zeta = 10^{-4}$. 
We observe that the difference is significantly smaller than the modulus of the QNM frequency in general relativity (unmodified), indicating that $\omega^{(1)}$ computed by the two approaches are well consistent. 
Specifically, the modulus difference is larger than $\zeta$ (the horizontal dashed line) but smaller than the dimensionless spin $a$ (the horizontal dashed-dotted line). 
This can be explained by the fact that $\omega^{(1)} (\rm MTE) $ is obtained by solving the modified Teukolsky equation which is correct up to the first order in $a$, whereas $\omega^{(1)} (\rm METRICS) $ is obtained using the full Kerr metric and metric corrections that are correct up to the second order in $a$. }
 \label{fig:Diff_w1_MET_vs_MTE_a_0005}
\end{figure}

We conclude this subsection by pointing out that, the validations using the Schwarzschild BH solution treat dCS gravity as an exact theory. 
However, we would like to remind the reader that dCS gravity is an effective field theory that serves as a lead-order approximation of more sophisticated theories. 
Thus, only the leading-order $\zeta$ modifications of the QNM frequencies and perturbations should be extracted.

\subsection{QNM frequencies of very-slowly rotating ($a = 0.005$) black holes }

We now extend our robustness check of the frequency of the axial perturbations of the 022, 032, and 033 modes to slowly rotating BHs with $a=0.005$.
We choose this value of $a$ because it is small enough to be accurately compared with calculations using the modified Teukolsky formalism and the eigenvalue perturbation method to linear order in spin. 
We compute the frequency of the 022, 032 and 033 modes because these are the ones that are involved in the analysis of actual GW signals \cite{LIGOScientific:2020tif, LIGOScientific:2021sio, Gennari:2023gmx}. 
We compute the METRICS frequency using the metric corrections and scalar field that satisfy the field equations up to second order in $a$. 
Figure~\ref{fig:Diff_w1_MET_vs_MTE_a_0005} shows the modulus of the difference between the $\omega^{(1)}$ computed using METRICS via Eq.~\eqref{eq:FWD} and that computed using the modified Teukolsky formalism. 
Observe that the modulus difference is small. 
Specifically, the modulus difference is significantly smaller than the modulus of the GR QNM frequencies. 
This small difference indicates a close consistency between the frequencies computed using the METRICS and the modified Teukolsky approaches. 

We also observe that the modulus difference is larger than $\zeta = 10^{-4}$ (horizontal dashed line), and is close to (but still smaller than) $a=0.005$ (horizontal dotted line). 
This indicates that the discrepancy between $\omega^{(1)}(\rm METRICS)$ and $\omega^{(1)}(\rm MTE)$ is dominated by the spin truncation. 
This can be explained by the fact that the modified Teukolsky computations are correct only up to the first order in $a$. 
In contrast, the METRICS computations use the full GR Kerr metric and the metric corrections and scalar field up to second order in $a$. 
Observe also that the modulus difference is significantly smaller than $a = 0.005$, which is roughly of $\mathcal{O} (a^2)$. 
The results of this check, together with the results of the $a=0$ subsection, indicate the close consistency between the results obtained with METRICS and those obtained with the modified Teukolsky formalism, thus validating the former in dCS gravity. 

\subsection{QNM frequencies of slowly rotating ($a = 0.1$) black holes}
\label{sec:a_01}

\begin{figure*}[htp!]
\centering  
\subfloat{\includegraphics[width=0.47\linewidth]{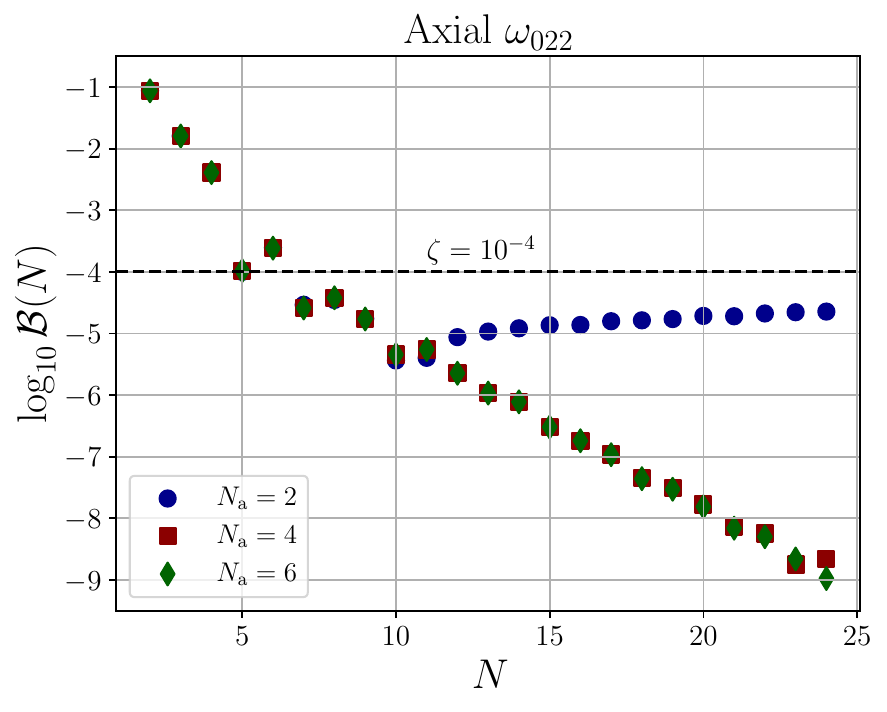}}
\subfloat{\includegraphics[width=0.47\linewidth]{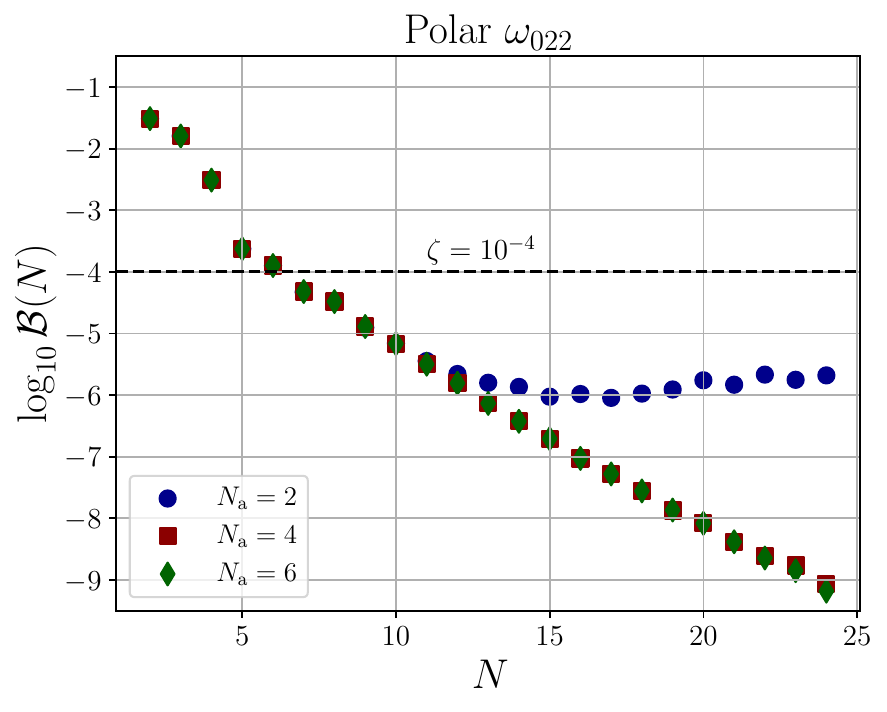}}
\caption{The backward modulus difference of the 022-mode frequency for axial (left) and polar (right) perturbations of a rotating BH with $a=0.1$ in shift-symmetric dCS gravity, with $\zeta = 10^{-4}$ (represented by the horizontal black dashed line), is shown as a function of the spectral order ($N$).
The background metric of the rotating BH incorporates modifications due to shift-symmetric dCS gravity up to $N_{\rm a} = 2$, 4, and 6 orders in $a$.
We find that for all $N_{\rm a}$, $\mathcal{B}(N)$ initially decreases exponentially before stabilizing and fluctuating around a constant value beyond a certain spectral order.
However, the level at which $\mathcal{B}(N)$ stabilizes can be significantly lowered by including higher-order metric corrections in $a$ from dCS gravity.
We have observed these features when we applied METRICS to EsGB gravity in \cite{Chung:2024vaf}, and their underlying cause is well understood (see main text and \cite{Chung:2024vaf} for details).
}
\label{fig:BWD_a_01}
\end{figure*}

As we extend our calculations to rotating BHs of $a = 0.1$, we observe an additional feature that also emerges in the computations for larger spin, as was also the case in EsGB gravity.
Figure~\ref{fig:BWD_a_01} shows $\mathcal{B}(N)$ of the 022-mode frequency of the axial (left) and polar (right) perturbations of rotating BHs with $a=0.1$ as a function of the spectral order $N$. 
The horizontal dashed line in Fig.~\ref{fig:BWD_a_01} shown when $\zeta = 10^{-4}$. 
What we are interested in studying is the backward modulus difference of $\omega^{(1)}$, which we can obtain from the backward modulus difference of $\omega$ through Eq.~\eqref{eq:FWD} because $|\omega(N)-\omega (N-1)| = \zeta |\omega^{(1)}(N)-\omega^{(1)}(N-1)|$. Therefore, the points below the horizontal dashed line have a backward modulus difference of $\omega^{(1)}$ smaller than 1. 
The frequencies are computed using the background metric and scalar field of a rotating BH that includes modifications due to the shift-symmetric dCS coupling up to $N_{\rm a} = 2$ (blue circles), 4 (red squares), and 6 orders (green diamonds) in $a$.

Observe that, for both parities and when $N_{\rm a} = 2$, $\mathcal{B}(N)$ first decreases approximately exponentially, and then reaches an approximate \textit{plateau} when $N\sim 10$.
We observed a similar feature when we applied METRICS to shift-symmetric EsGB gravity \cite{Chung:2024ira, Chung:2024vaf}, and found that this phenomenon stems from the fact that the background metric and the scalar-field profile of the rotating BHs satisfy the field equations only up to a certain order in $a$. 
We thus expect that the approximate nature of the background metric and the scalar field is also the cause of this phenomenon in dCS gravity. 
To validate this hypothesis, we examine $\mathcal{B}(N)$ when $N_{\rm a} = 4 $ and 6. 
When $N_{\rm a} = 4$, the minimal $\mathcal{B}(N)$ is significantly smaller than that when $N_{\rm a} = 2$, and is attained at $N \sim 24 $ (the axial mode), a significantly larger value than the spectral order that minimizes $\mathcal{B}(N)$ when $N_{\rm a} = 2 $.
When $N_{\rm a} = 6$, we observe no plateau in both parities for the values of $N_{\rm a}$ explored. 
The tendency that minimal $\mathcal{B}(N)$ decreases as $N_{\rm a}$ increases proves our hypothesis that the error in the background metric and the scalar field is the cause of the plateaus. 

This observation guides us in choosing a sufficiently large $N_{\rm a}$ such that the background metric and scalar field are accurate enough for the computation of the QNM frequencies at a given $a$.
In Sec.~VI B of \cite{Chung:2024vaf}, we estimated that the minimal $\mathcal{B}(N)$ of $\omega^{(1)}$ should be smaller than $10^{-4}$ so that it is accurate enough to analyze existing and future ringdown signals. 
Given Fig.~\ref{fig:BWD_a_01}, we then notice that we need $a^{N_{\rm a}} \leq 10^{-4} $ to achieve this desired minimal $\mathcal{B}(N)$ at a given $a$, as in the case of EsGB gravity. 
That is, we need to select the smallest integer $N_{\rm a}$ such that $a^{N_{\rm a}} \leq 10^{-4} $ for a given choice of $a$. 
After selecting $N_{\rm a}$, we extract the QNM frequency of a given mode, parity, $a$, and $\zeta$ via the following procedure: 
\begin{enumerate}
    \item We compute the QNM frequencies for $N \leq 25$. 
    We terminate the calculations at $N = 25$ because we find that $\mathcal{B}(N)$ is usually saturated or minimized when $20 \leq N \leq 25 $, as in the case of EsGB gravity \cite{Chung:2024ira, Chung:2024vaf}. 
    \item We select the optimal spectral order, $N_{\rm opt}$, via the following criterion
    \begin{equation}\label{eq:opt_spec_order}
    N_{\rm opt} = \text{arg}\min_{N} \mathcal{B}(N). 
    \end{equation}
    \item We select the frequency at the optimal spectral order, 
    \begin{equation}\label{eq:opt_QNMF}
    \omega_{\rm opt} = \omega(N_{\rm opt}), 
    \end{equation}
    as the frequency at that QNM, parity, $a$ and $\zeta$. 
    \item We interpret the modulus of the minimal backward displacement, 
    \begin{equation}\label{eq:numerical_uncert}
    \text{MBD} = \omega(N_{\rm opt}) - \omega(N_{\rm opt} - 1), 
    \end{equation}
    as the numerical uncertainty of both the real and imaginary parts of the QNM frequencies. 
\end{enumerate}

\section{Numerical Results for rotating black holes with moderate spins}
\label{sec:Results-2}

We here present the results obtained by solving the linearized, algebraic equations with the Newton-Raphson method for rotating black holes with moderate spin. We first describe our numerical results, and then we develop fitting functions that can represent the QNM frequency shifts in dCS smoothly in spin. 

\subsection{QNM frequencies of black holes with moderate rotation}
\label{sec:Intermediate_spinning}

\begin{figure*}[htp!]
\centering  
\subfloat{\includegraphics[width=0.47\linewidth]{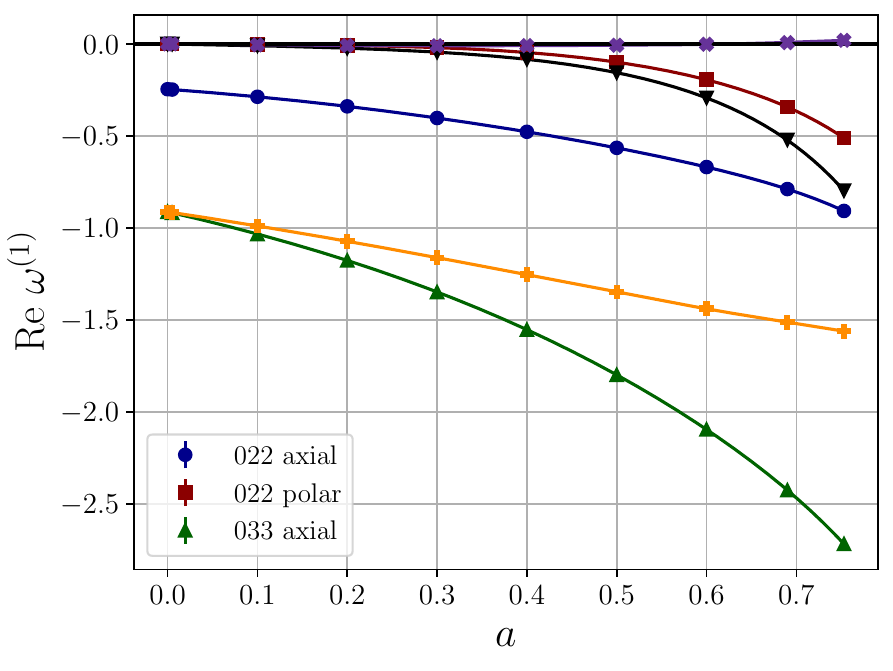}}
\subfloat{\includegraphics[width=0.47\linewidth]{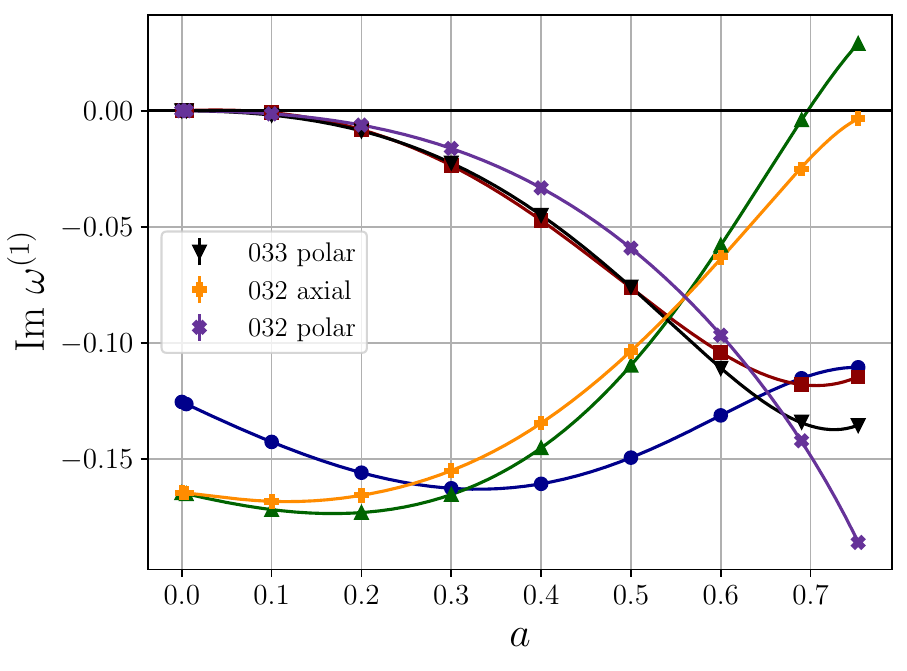}}
\caption{The real (left) and imaginary (right) parts of $\omega^{(1)}$ for the axial (blue dots) and polar (red triangles) metric perturbations of the $nlm = 022$, 033, and 032 modes are shown as functions of the dimensionless spin.
The solid color lines represent the $\omega^{(1)}$ computed using the optimal fitting polynomial of the corresponding mode. 
We observe that $\omega^{(1)}$ differs between axial and polar perturbations, confirming that isospectrality is broken in dCS gravity.
Specifically, we observe that the magnitude of $\omega^{(1)}$ is smaller for polar perturbations than for axial perturbations, and for $a \lesssim 0.3$, both $\text{Re}\;\omega^{(1)}$ and $\text{Im}\;\omega^{(1)}$ exhibit distinct behaviors for axial perturbations.
These observations align with prior studies on slowly rotating BHs in dCS gravity.
Lastly, although we have included the numerical uncertainty of our frequency computations, the standard deviation of $\omega^{(1)}$ as we fit Eq.~\eqref{eq:w_zeta} using the METRICS frequencies, as error bars, the error bars for the considered range of $a$ are too small to be visible in the figure and are instead shown in Fig.\ref{fig:delta}.
The small uncertainties further manifest the accuracy of our computed $\omega^{(1)}$.
}
\label{fig:omega_1_022}
\end{figure*}

We perform the aforementioned procedures to compute the QNM frequency of rotating BHs with larger $a$. 
At different $a$, we aim to estimate $\omega^{(1)}$. 
Although the forward difference scheme [Eq.~\eqref{eq:FWD}] can quickly validate the METRICS frequencies, the scheme does not adequately address the following sources of errors. 
\begin{enumerate}
    \item The error of METRICS in computing the QNM frequencies of Kerr BHs in GR. 
    In \cite{Chung:2023wkd}, we find that the error in the 022-mode METRICS frequency can be as large as $\sim 10^{-6}$ for $ 0.5 \leq a \leq 0.8 $. We wish to estimate the dCS shift to the frequencies as accurately as possible, and to avoid the error in the GR Kerr frequencies to contaminate the shift. For this reason, we calculate $\omega^{(0)}$ using METRICS [i.e.~$\omega^{(0)} = \omega(\zeta = 0|\rm METRICS)$] and then subtract this quantity from $\omega(\zeta|\rm METRICS)$ using METRICS [i.e.~$\zeta \omega^{(1)} = \omega(\zeta|\rm METRICS) - \omega(\zeta = 0|\rm METRICS)$]. Since the error in the GR Kerr frequencies is not random (but rather due to the approximate nature of the asymptotic factor~\cite{Chung:2023wkd}), this error cancels out when computing the shift in this way.   
    \item The error due to the truncation of the background metric modifications and background scalar field to a given order in $a$ ($N_{\rm a}$). 
    We keep this error controlled by choosing a sufficiently large $N_{\rm a}$ such that $a^{N_{\rm a}} \leq 10^{-4}$, guided by the observations in Sec.~\ref{sec:a_01}. 
    Figure~\ref{fig:BWD_a_01} shows that this choice of $N_{\rm a}$ is sufficient to keep this error small. 
    
    To improve the accuracy of METRICS at higher spin values, we are currently developing a method for constructing the background metric and scalar field of very rapidly spinning black holes in quadratic gravity, including in dCS gravity, which will be presented in a forthcoming publication. These improved backgrounds will enable accurate QNM computations for very-rapidly rotating black holes using METRICS, as will be described in a subsequent study.

    \item The error due to the truncation of the spectral expansions of the perturbations to a specific spectral order ($N$). 
    To keep this error under control, we select the optimal spectral order according to Eq.~\eqref{eq:opt_spec_order}, such that $\mathcal{B}(N)$ is minimized. 
    \item The nonlinear dependence of the QNM frequencies on $\zeta$ (even $\zeta \ll 1 $) at a given $a$. 
    More generally, we expect the QNM frequency to depend on $\zeta$ via 
    \begin{equation}
    \omega = \omega^{(0)} + \omega^{(1)} \zeta + \omega^{(2)} \zeta^2 + \omega^{(3)} \zeta^3 + {\cal{O}}(\zeta^4). 
    \end{equation}
    Ignoring the terms of quadratic and higher-order in $\zeta$ will also bias our estimate of $\omega^{(1)}$. 
    To reduce this error, we choose a sufficiently small $\zeta$ to compute the linear shift.
    In this work, except when $a=0$, we only consider $\zeta \leq 10^{-4}$ for METRICS calculations, which imply an error of ${\cal{O}}(10^{-8})$ or smaller. 
    \item The error due to the truncation of the Lagrangian up to quadratic terms in the Riemannian curvature tensor. 
    This error is for sure subdominant as long as $\zeta \ll 1$, but precisely how large can $\zeta$ be before we exit the regime of validity of the effective field theory requires further study.    
\end{enumerate}

Errors like these will always be present in QNM calculations beyond GR, but we have made a conscientious effort in this paper to minimize them as much as possible and to estimate them to quantify the uncertainty in our calculations. 
We need to consider all of these errors to estimate the combined error of $\omega^{(1)}$. 
To this end, we first use METRICS to estimate $\omega$ for a given, fixed value of $a$ (according to the procedures in Sec.~\ref{sec:a_01}) at three different values of $\zeta$ that are close to each other;
we choose these values to be $\zeta = 0, 5 \times 10^{-5}$ and $10^{-4}$, so that the small-coupling approximation applies. 
Then, using the METRICS frequencies [Eq.~\eqref{eq:opt_QNMF}] at these three $\zeta$ as data, together with their corresponding numerical uncertainty [Eq.~\eqref{eq:numerical_uncert}], we estimate $\omega^{(1)}$ by fitting Eq.~\eqref{eq:w_zeta} as a function of $\zeta$ using the Python \texttt{curvefit} function. 
The best-fit value and the standard deviation ($\sigma$) are finally taken as the METRICS $\omega^{(1)}$ and its numerical uncertainty, respectively, for the given value of $a$. 
In particular, this numerical uncertainty combines the two largest sources of error in our calculation, i.e.~items 1 and 4 in the above list. 

Figure~\ref{fig:omega_1_022} shows the best-fit real and imaginary parts of $\omega^{(1)}$ of the 022-, 033- and 032-mode frequencies of the axial and polar perturbations, with their numerical certainty included as error bars. 
Tables~\eqref{tab:omega_1_022},~\eqref{tab:omega_1_033}, and~\eqref{tab:omega_1_021} of Appendix~\ref{sec:Appendix_B} show the value of the best-fit real and imaginary parts of $\omega^{(1)}$ of the frequency of these three modes for both parities. 
We estimate $\omega^{(1)}$ and its error from $a=0$ to $a \sim 0.75 $ at steps of $\sim 0.1$.
We terminate our computations at $a \sim 0.75$ because we find that the background dCS metric and the background scalar field at $40$ th order in $a$ are not accurate enough for METRICS computations of QNM frequencies in dCS gravity when the spin is above this value. 
To better visualize the variation of $\omega^{(1)}$ over $a$, the value of $\omega^{(1)}$ computed using the optimal fitting polynomial of the corresponding mode (see Sec.~\ref{sec:fitting_exp}) is shown with solid color lines in Fig.~\ref{fig:omega_1_022}. 

This figure and tables allow us to make several observations. 
First, observe that, at $a=0$, the frequency of the 032 and 033 modes are the same, because the QNM frequencies of a nonrotating BH are independent of the magnetic mode number $m$; this feature serves as a sanity check of our computations. 
Second, observe that isospectrality is not preserved in dCS gravity because $\omega^{(1)}$ is different for different parity modes. 
This is a feature that emerges in various modified gravity theories \cite{Manfredi:2017xcv, Chen:2021cts, QNM_dCS_01, QNM_dCS_02, QNM_dCS_03, QNM_dCS_04, QNM_EdGB_01, QNM_EdGB_02, QNM_EdGB_03}, and that has been found to be generic in modified gravity~\cite{Li:2023ulk}.
Explicitly, we observe that $|\omega^{(1)}|$ is larger for axial perturbations than for polar perturbations, which can be explained by the fact that the axial perturbations couple more strongly to the dCS terms. 

\begin{figure}[tp!]
\centering  
\subfloat{\includegraphics[width=\columnwidth]{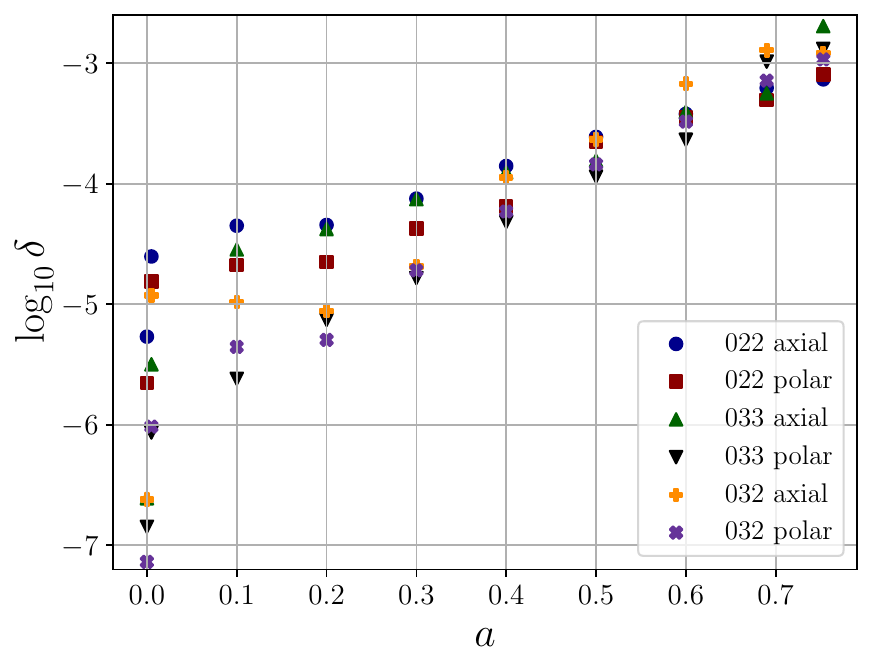}}
\caption{
The numerical uncertainty, quantified as the modulus of the standard deviation when fitting Eq.~\eqref{eq:w_zeta} to estimate $\omega^{(1)}$, is shown as a function of the dimensionless spin.
We generally observe that the uncertainty grows with increasing $a$.
This trend arises because the metric corrections and the background scalar field satisfy the field equations only up to a finite order in the dimensionless spin.
Although we have adjusted the spin-truncation order of the background metric retained for QNM frequency computations, the error in the background still increases as $a$ grows.
Nevertheless, across all computed values of the dimensionless spin, $\delta$ remains significantly smaller than the typical magnitude of QNM frequencies in GR, which is approximately $\mathcal{O}(10^{-1})$.
Thus, the METRICS-derived $\omega^{(1)}$ values are sufficiently accurate for analyzing GW data.
}
\label{fig:delta}
\end{figure}

Beyond isospectrality breaking, we also observe in Fig.~\ref{fig:omega_1_022} that the error bars representing numerical uncertainty are too small to be seen, while keeping the legends a reasonable size; this is a clear, visual manifestation of the accuracy of the METRICS computations of $\omega^{(1)}$. 
Since the numerical uncertainty is too small to be seen in the figure, we visualize it separately in Fig.~\ref{fig:delta} as a function of $a$. 
Observe that the combined error in the QNM frequencies increases with $a$ in general, which we believe is a feature of our background metric and scalar field becoming less accurate with increasing spin. 
In spite of this trend, the QNM frequencies computed at all $a$ are still well within the chosen accuracy ($10^{-3}$); we recall that this accuracy threshold was estimated in \cite{Chung:2024ira} to ensure that QNM frequencies are accurate enough to analyze future ringdown signals that will be detected by next-generation ground-based detectors. 
Thus, our results are accurate enough to test dCS gravity with existing and future ringdown signals. 

Finally, we conclude this subsection with a minor remark about the accuracy of the Newton-Raphson method to solve the linearized algebraic equations. At any point in the Newton-Raphson iteration scheme, we can compute a ``residual error'' by evaluating the linearized algebraic equations at that iteration point; the residual error will typically be a vector, so we take its $L^2$ norm to obtain a number. With this in hand, we can then compute the residual error ratio as the ratio between the residual error at the initial guess and at the final iteration point (defined by Eq. (52) of \cite{Chung:2023wkd}). We have checked that this residual error ratio for all modes we investigated is of $\mathcal{O}(10^{-8}) $ for all $a$ we computed. 
Such a small residual error ratio indicates that the numerical solutions obtained by the Newton-Raphson algorithms satisfy the linearized field equations very accurately. The linearized equations themselves, however, are not exact, because of the reasons explained in the list above. This is why we are careful to estimate our uncertainties in this subsection, which are not dominated by our Newton-Raphson numerical scheme, but rather, by the approximate nature of the background dCS metric and scalar field.  

\subsection{Fitting function for the leading-order-in-$\zeta$ QNM frequency shift for moderately spinning black holes}
\label{sec:fitting_exp}

\begin{figure*}[htp!]
\centering  
\subfloat{\includegraphics[width=0.47\linewidth]{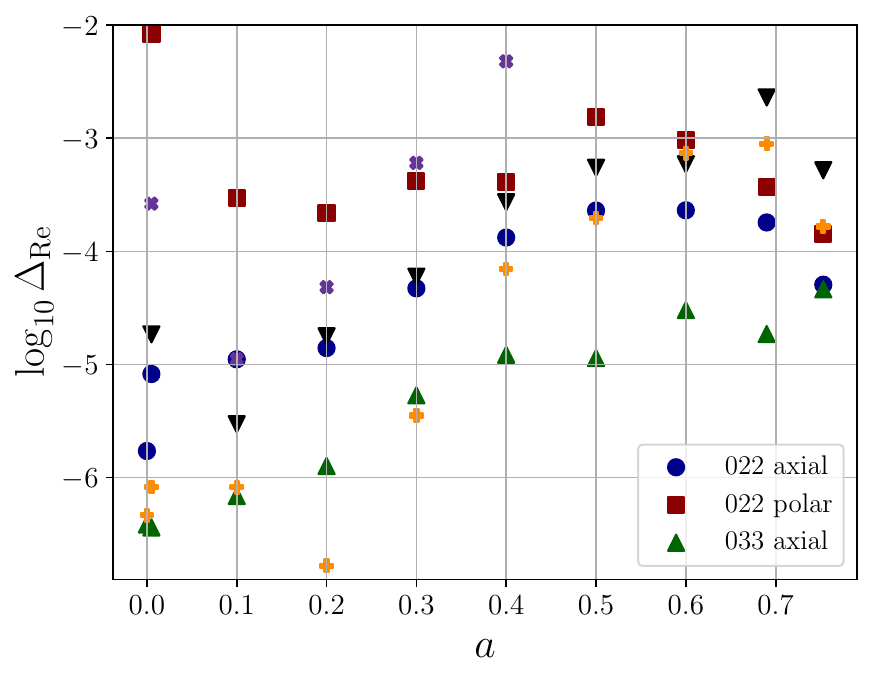}}
\subfloat{\includegraphics[width=0.47\linewidth]{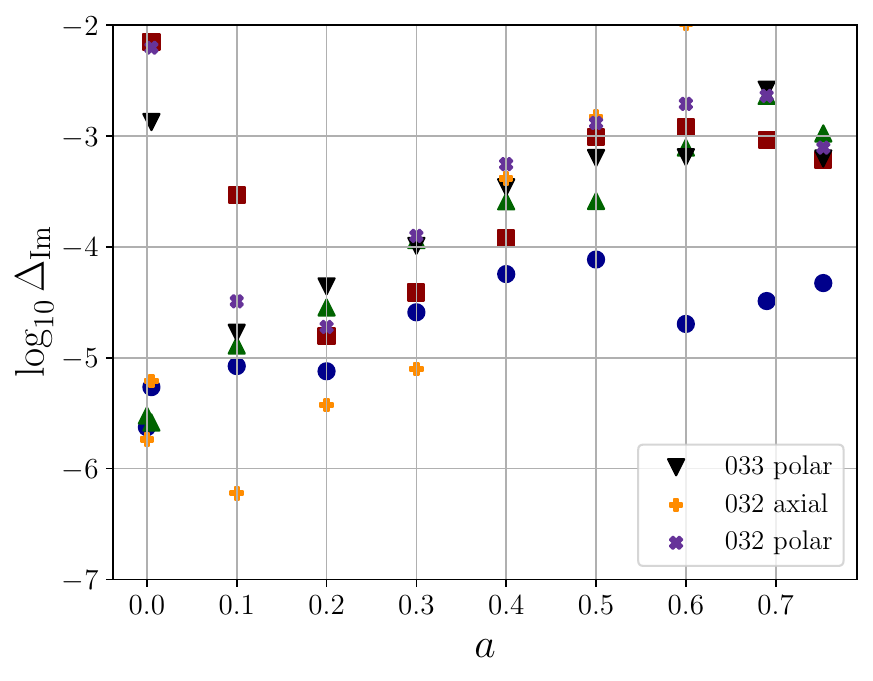}}
\caption{The relative fractional errors in the real (left) and imaginary (right) parts of $\omega^{(1)}$, comparing results from the optimal fitting polynomials with those computed using METRICS. 
We observe that the relative fractional error $< 10^{-2}$, indicating that the optimal fitting polynomials provide highly accurate, yet rapid, computations of the QNM frequencies in dCS gravity. 
}
\label{fig:fitting_polynomials_check}
\end{figure*}

Ringdown tests of dCS gravity require evaluating the QNM frequencies in this theory at a vast number of sampling points, e.g.~over many values of spin and coupling constant. 
It is impractical to solve the linearized field equations at every one of these sampling points ``on the fly'', because the computation cost is extremely high. 
To make ringdown tests a reality, we must construct an \textit{emulator} of the dCS QNM frequency shift data, which we here achieve with polynomial fitting functions. We will show below that such an emulator is accurate enough for our purposes and sufficiently fast for use in parameter estimation. 

Let us then fit the real and imaginary parts of the METRICS dCS frequency shifts with polynomial of the form 
\begin{equation}\label{eq:omega_1_fitted}
\begin{split}
\omega^{(1)} & = \sum_{j}^{N_p} w_{j} a^j\,. 
\end{split}
\end{equation}
where we have set $M=1$, because the black hole mass scales out, while $w_j$ and $N_p$ are complex constants and the polynomial order respectively, both of which are to be determined by the fitting algorithm. 
The lower limit of the summation depends on the parity of the perturbations. 
For polar-sector perturbations, $w_0 = 0 $ follows from the property that $\omega^{(1)}(a=0|\zeta) = 0$, as the polar-sector perturbations and dCS coupling terms decouple in the Schwarzschild limit.
Thus, the $j$ sum starts at 0 for axial perturbations, and 1 from polar perturbations.
For a given value of $N_p$, a given $(n l m)$ mode and a given parity, we use the built-in \texttt{NonLinearModelFit} function in \textit{Mathematica} to determine $w_j$ and the $1\sigma$ fitting uncertainty in these coefficients, using the METRICS QNM frequency shifts and their uncertainties as data. Obviously, as $N_p$ increases, the polynomial is able to fit the data better and better, but if $N_p$ is too large, overfitting occurs and spurious oscillations are introduced. We pick $N_p$ such that the fit minimizes the loss function
\begin{equation}
\begin{split}
& {\rm{Loss}} (N_p) \\
& = \frac{1}{N-N_p-1}\sum_{k=1}^{N}\left| \omega^{(1)} (a_k|\text{METRICS}) - \sum_{j}^{N_p} w_j a_{k}{}^j\right|^2, 
\end{split}
\end{equation}
where $a_k$ for $k=1, 2, ..., N$ is a grid points in $a$ at which we computed the QNM frequencies with METRICS, and $N$ is the number of grid points on the dimensionless-spin line. 
Note that the number of parameters in the fitting function are included in the loss function, so that by minimizing it, we are attempting to prevent overfitting. 

We list the explicit numerical value and uncertainty of the best-fit $w_j$ for the optimal polynomial fit in Tables~\ref{tab:poly_fit_coeffs} and \ref{tab:poly_fit_coeffs_uncertainty}. 
Below we provide the fitting polynomials truncated at $a^4$ for quick reference 
\begin{widetext}
\begin{align}
\omega_{022,\rm A} = & \left[-0.246041 - 0.125482i \pm (4.80 \times 10^{-6} + 4.22 \times 10^{-7} i)\right] 
+ \left[ -0.35908-0.184144 i \pm (0.00513 + 0.000404i) \right] a 
\nonumber \\
& + \left[ -0.595672 + 0.0876598i \pm (0.135 + 0.00915i) \right] a^2 
+ \left[1.41373 + 0.258192i \pm (1.45 + 0.0777i) \right] a^3 
\nonumber \\ 
& + \left[ -11.2668 + 0.703524i \pm (8.17 + 0.321i) \right] a^4 + \ldots\,, \\
\omega_{022,\rm P} = & \left[-0.0307839 + 0.017411i \pm (0.000224 + 0.00145i) \right] a 
+ \left[0.013835 - 0.214353i \pm (0.0644 + 0.0335i) \right] a^2 
\nonumber \\
& + \left[0.398761 - 0.447376i \pm (0.692 + 0.0288i) \right]a^3 
+ \left[ -6.66087 + 0.448166i  \pm (3.93 + 1.21i) \right] a^4 + \ldots\,, \\
\label{eq:METRICS_fitting_polynomials}
\omega_{033,\rm A} = & \left[-0.912752 - 0.164615i \pm (7.29 \times 10^{-8} + 8.18 \times 10^{-8} i) \right] 
+ \left[-1.09512 - 0.0911042i \pm (0.000221 + 0.000239i) \right] a 
\nonumber \\
& + \left[-1.2327 + 0.116646i \pm (0.0131 + 0.00837i) \right] a^2 + \left[2.37515 + 0.763688i \pm (0.235 + 0.105i) \right] a^3 
\nonumber \\
& + \left[-16.7427 - 0.927771i \pm (1.75 + 0.566i) \right]a^4 + \ldots\,, 
\\
\omega_{033,\rm P} = & \left[ - 0.0901535 - 0.000837525i  \pm (0.000245 + 0.0000810i) \right] a 
+ \left[ -0.0763526 - 0.149877i \pm (0.0177 + 0.00251i) \right] a^2 
\nonumber \\
& + \left[ 0.452064 - 0.248466i \pm (0.363 + 0.0343i) \right] a^3
+ \left[ -6.70664 - 0.508568i \pm (2.98 + 0.211i) \right] a^4 + \ldots\,, \\
\omega_{032,\rm A} = & \left[ -0.912726 - 0.164606i \pm (5.67 \times 10^{-7} + 3.43 \times 10^{-7}i ) \right]
+ \left[ -0.721554 - 0.0266668i  \pm (0.00637 + 0.00385i) \right] a 
\nonumber \\
& + \left[ -0.815162 - 0.717186i  \pm (0.173 + 0.104i) \right] a^2 
+ \left[ 6.14597 + 9.84411i  \pm (2.03 + 1.23i) \right] a^3 
\nonumber \\
& + \left[ -38.4939 - 49.307i \pm (13.1 + 7.90i) \right] a^4 + \ldots\,, \\
\omega_{032,\rm P} = & \left[ 0.0213119 + 0.000953349i  \pm (0.000416 + 0.000160i) \right]a 
+\left[ -1.98616 - 0.154152i \pm (0.0284 + 0.0109i) \right] a^2 
\nonumber \\
& +\left[ 20.3576 + 0.42749i \pm (0.572 + 0.220i) \right] a^3 
+ \left[ -101.368 - 4.25973i \pm (4.67 + 1.79i) \right] a^4 + \ldots\,,
\end{align}
\end{widetext}
The complex number in parentheses following the $\pm$ sign represents the $1\sigma$ uncertainty of the corresponding coefficient to three significant digits. The full fits are to higher order in $a$, but we do not present them here explicitly, and instead list the fitting coefficients in Tables~\ref{tab:poly_fit_coeffs} and \ref{tab:poly_fit_coeffs_uncertainty}.

To assess the accuracy of the optimal fitting polynomials in computing $\omega^{(1)}$, we compare their estimates with those obtained directly from METRICS.  
Figure~\ref{fig:omega_1_022} shows the real (left) and imaginary (right) parts of $\omega^{(1)}$ computed using the optimal fitting polynomial of the corresponding modes in the range $0 \leq a \leq 0.753$ with solid color lines, and the $\omega^{(1)}$ computed using METRICS with symbols.
All curves pass through the METRICS frequencies of the corresponding mode almost perfectly, as expected. 
The close alignment between the curves and the scattered points provides a visual confirmation of the fitting accuracy.  

For a more quantitative assessment, Fig.~\ref{fig:fitting_polynomials_check} shows the relative fractional errors, defined as
\begin{equation}
\Delta_{\rm Re / Im} = \frac{\omega_{\rm Re/Im}^{(1)} (a|\text{METRICS}) - \sum_{j}^{N_p} w_j a_{k}{}^j}{\omega_{\rm Re/Im}^{(1)}  (a|\text{METRICS})}, 
\end{equation}
for both the real and imaginary parts of $\omega^{(1)}$ calculated using the optimal fitting polynomial.  
Across the range considered, the errors remain below $\lesssim 10^{-2}$, demonstrating that the optimal fitting polynomials provide highly accurate, yet rapid, computations of the METRICS frequencies in dCS gravity.  
This level of precision suggests that the optimal fitting polynomials are well-suited for applications in GW data analysis. 

\begin{figure*}[htb!]
\centering  
\subfloat{\includegraphics[width=6cm]{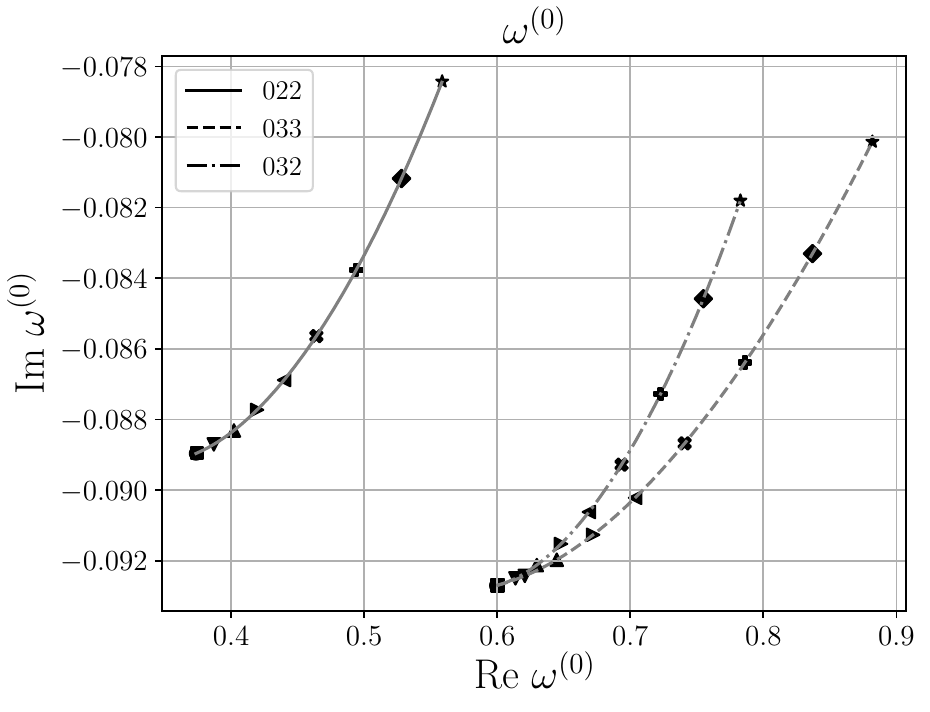}}
\subfloat{\includegraphics[width=6cm]{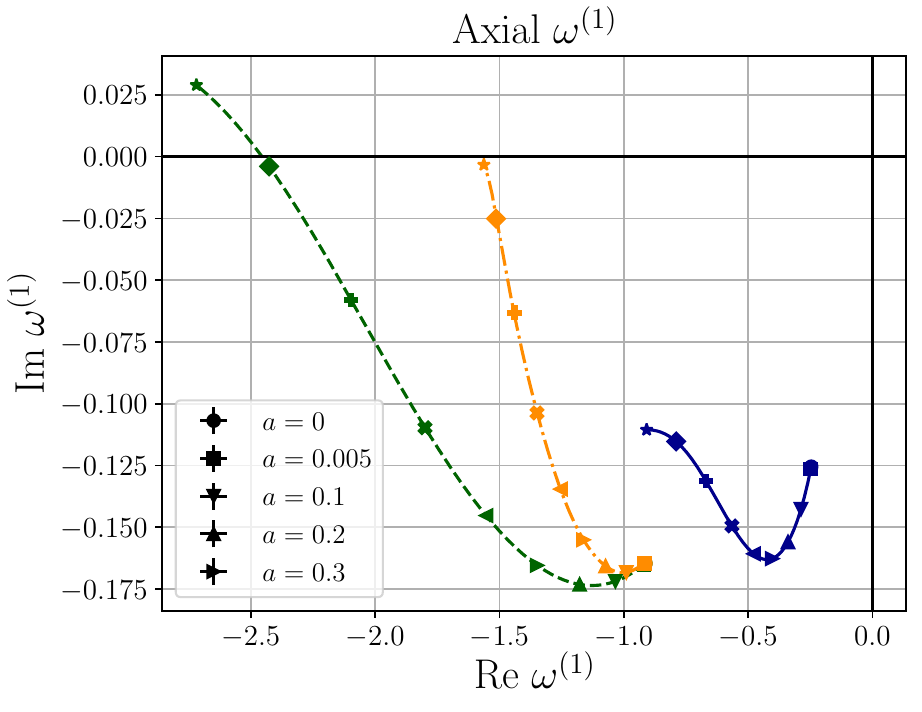}}
\subfloat{\includegraphics[width=6cm]{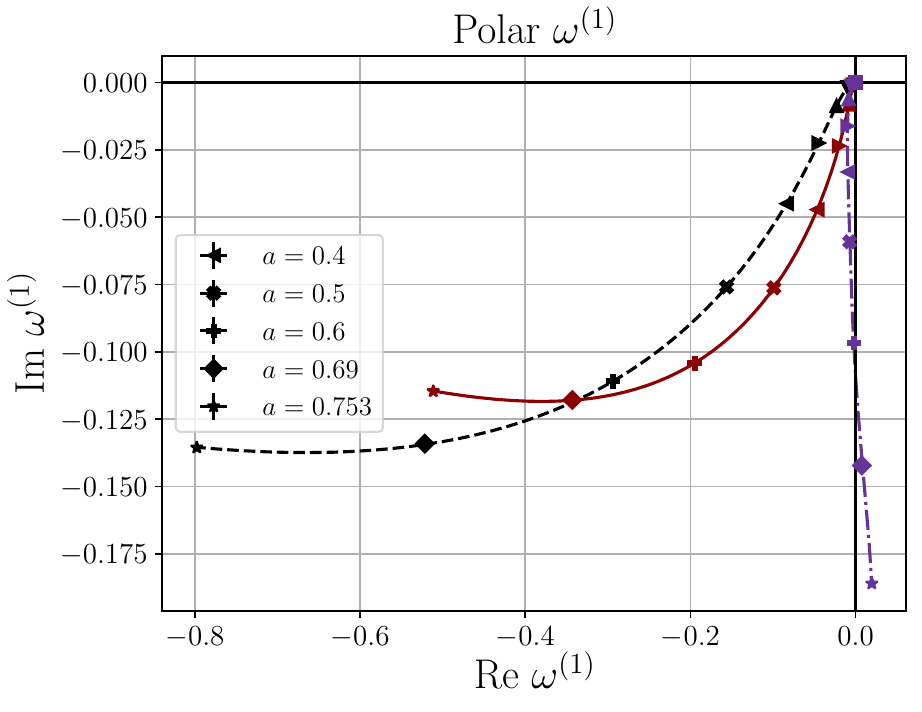}}
\caption{Complex trajectories of $\omega^{(1)}$ for the axial (middle) and polar (right) perturbations, and of the $022$ (solid), $033$ (dashed), and $032$ (dashed-dotted) modes as $a$ increases from $0$ to $0.753$. Scattered points represent METRICS frequencies from $a=0$ (circle) to $a=0.753$ (star). 
The real and imaginary axes are marked by solid black lines. 
The left panel shows the corresponding trajectories of $\omega^{(0)}$ in GR computed from the Teukolsky equation, and the METRICS GR frequencies represented by scatters. Observe how the dCS shifts in the QNM frequency are not linear functions of spin. 
}
\label{fig:complex_planes}
\end{figure*}

The optimal fitting polynomials provide a clearer visualization of the leading-$\zeta$-order shift in QNM frequencies in the complex plane.  
The middle and right panels of Fig.~\ref{fig:complex_planes} illustrate the complex trajectories of $\omega^{(1)}$ traced by the optimal fitting polynomials for the axial (middle) and polar (right) perturbations of the $022$ (solid line), $033$ (dashed line), and $032$ (dashed-dotted line) modes as $a$ increases from $0$ to $0.753$.  
Scatters mark the METRICS frequencies, ranging from $a=0$ (circle) to $a=0.753$ (star).  
The real and imaginary axes are indicated by solid horizontal and vertical black lines, respectively.  
We observe that all polar-mode trajectories originate from the origin, which aligns with the expectation that, for $a=0$, polar-sector metric perturbations are decoupled from the dCS terms, leaving the polar-mode frequencies unmodified.  
For comparison, the left panel of Fig.~\ref{fig:complex_planes} presents the complex trajectories of $\omega^{(0)}$ for the $022$ (solid gray line), $033$ (dashed gray line), and $032$ (dashed-dotted gray line) modes, computed using the Teukolsky equation for $a$ ranging from $0$ to $0.753$.  

\section{Implications to gravitational-wave data analysis} 
\label{sec:DA}

\begin{figure}[htp!]
\centering  
\subfloat{\includegraphics[width=\columnwidth]{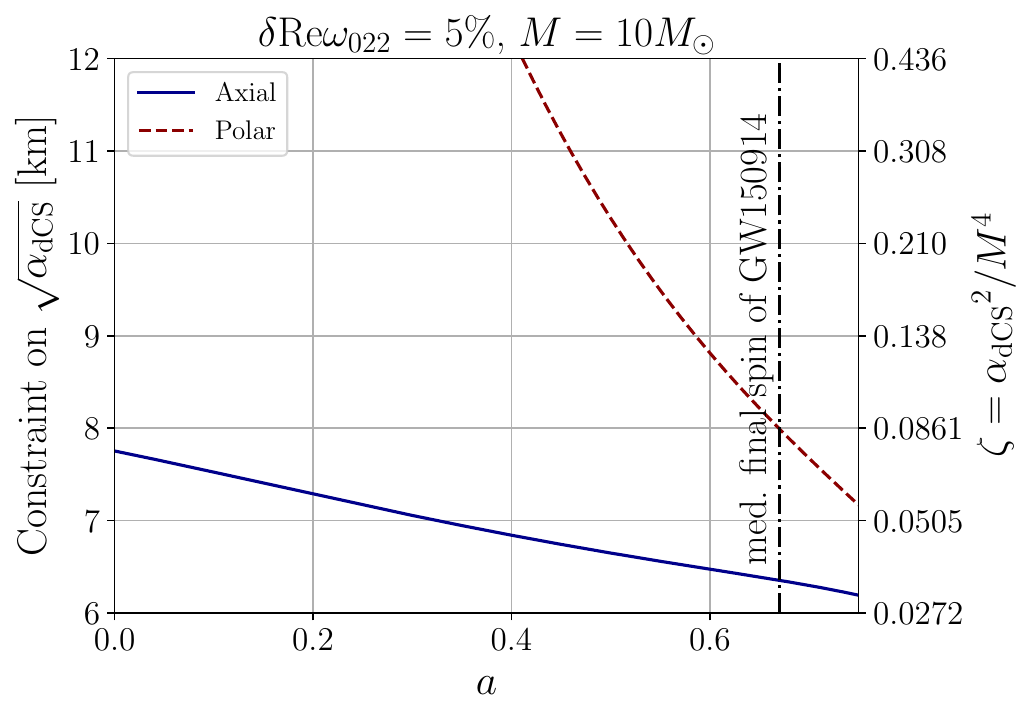}}
\caption{Projected constraints on the square root of dCS coupling constant, $\sqrt{\alpha_{\rm dCS}}$ (left vertical axis), derived from a hypothetical 5$\%$ relative measurement uncertainty in a future estimation of the 022-mode real frequency via Eq.~\eqref{eq:constraint}. 
In this figure, we assume the remnant BH has a mass of $10 M_{\odot}$ for ease of rescaling.
The right vertical axis shows the value of the dimensionless coupling parameter, $\zeta = \alpha_{\rm dCS}^2 / M^4$ for the corresponding value of $\alpha_{\rm dCS}$. 
Observe that constraints with both axial and polar frequencies become more stringent as the remnant BH spin $a$ increases, because then the dCS shifts of the QNM frequencies become greater. 
For $a \lesssim 0.4$, the constraint obtained from the polar frequency is significantly larger than that from the axial frequency because the polar frequency of slowly rotating BHs is not significantly shifted. This forces the polar constraints to exist the regime of validity of the effective field theory, and thus, such polar shifts cannot be used to place a constraint.   
However, for $a\sim 0.7$, which corresponds to the most likely spin of many remnant BHs of GW signals detected by the LIGO-Virgo-KAGRA detectors (as an example, the median remnant spin of GW150914 is marked by a dashed vertical line), the projected constraints on $\sqrt{\alpha_{\rm dCS}}$ correspond to $\zeta \lesssim 0.1$, which is well within the small coupling regime. 
This suggests that our computed QNM spectra for dCS black holes have strong potential for establishing the first meaningful constraints on the dCS coupling parameter with QNM ringdown observations in the small-coupling approximation. 
}
 \label{fig:constraints}
\end{figure}

The QNM spectra obtained in this work enable us to perform GW test of dCS gravity through ringdown detection.  
In this section, we estimate the constraints on $\sqrt{\alpha_{\rm dCS}}$ that can be derived if we detect ringdown signals in the future that happen to be consistent with GR.  
Specifically, we focus on estimating the constraint that can be placed using the real part of the 022-mode frequency, as this is the dominant mode in detected ringdown signals.  
Moreover, the relative measurement uncertainty of the real part is typically significantly smaller than that of the imaginary part.  
As a result, the real frequency of the 022 mode will contribute the most to the constraint on $\sqrt{\alpha_{\rm dCS}}$.

Isospectrality is not preserved in dCS gravity.  
If the detected ringdown signal is consistent with GR (which is an isospectral theory), only a single 022-mode frequency will be measured (because even and odd parity modes would have the same frequency).  
Using the dCS prediction for frequencies of different parities will lead to different constraints.  
To simplify the estimation, let us assume that the signal is known to be purely axial or purely polar. 
If the relative measurement uncertainty in the real part of the 022 mode is $ \delta \text{Re} ( \omega_{022} ) $ and if the signal is consistent with GR, then it must be that $ \zeta $ satisfies the inequality  
\begin{equation}
\zeta \frac{\omega^{\rm Re, (1)}_{022}}{\omega^{\rm Re, (0)}_{022}} \leq \delta \text{Re} ( \omega_{022} ). 
\end{equation}
The definition $ \zeta = \alpha_{\rm dCS}^2 / M^4 $ allows us to translate this inequality into a constraint on $ \sqrt{\alpha_{\rm dCS}} $:  
\begin{align}\label{eq:constraint}
\sqrt{\alpha_{\rm dCS}} \leq & 7.99 \;  {\rm km} 
\left( \frac{\delta \text{Re} ( \omega_{022} )}{0.05} \right)^{1/4} 
\left( \frac{M}{10 M_{\odot}} \right)
\nonumber \\
& \times \left( \frac{\omega^{\rm Re, (0)}_{022}/\omega^{\rm Re, (1)}_{022}}{0.0343}\right)^{1/4} 
\,.
\end{align}
Here $0.0343$ is approximately the ratio of $\omega^{\rm Re, (0)}_{022}$ to $\omega^{\rm Re, (1)}_{022}$ at $a=0.7$. 
As expected, this constraint on $ \sqrt{\alpha_{\rm dCS}} $ is proportional to $ M $, so the smaller the final black hole mass, the more stringent the constraint.
Moreover, unless $ a \sim 0 $, we have $ \text{Re} ( \omega_{022} ) \sim \omega^{\rm Re, (0)}_{022} \sim \omega^{\rm Re, (1)}_{022} \sim \mathcal{O}(0.1) $.  
Due to the quartic root, the proportionality factor $ \sim \mathcal{O}(1) $.  
This estimation suggests that the constraint on $ \sqrt{\alpha_{\rm dCS}} $ is of the order of a fraction of the event horizon size, which is reasonable given that we are constraining dCS gravity by measuring its effects on the pulsations of BH event horizon or its light ring. 

Figure~\ref{fig:constraints} presents plausible constraints on $\sqrt{\alpha_{\rm dCS}}$ that one can estimate with Eq.~\eqref{eq:constraint}, assuming $M = 10M_{\odot}$ and $\delta \text{Re} ( \omega_{022} ) = 5\%$. 
We consider a rotating BH with a mass of $10 M_{\odot}$ because the constraints obtained for this mass can be conveniently scaled to BHs of different masses. 
The choice of $\delta \text{Re} ( \omega_{022} ) = 5\%$ reflects the typical relative measurement uncertainty in the real part of the 022-mode frequencies in detected ringdown signals.
The right vertical axis of Fig.~\ref{fig:constraints} shows the corresponding value of the dimensionless dCS coupling parameter for each 
value of $\sqrt{\alpha_{\rm dCS}}$ on the left vertical axis.
We set the range of the vertical axis so that all calculations are within the cutoff scale of the effective field theory, i.e.~within $\zeta  \lesssim 1/2$.
For both the axial and polar 022 modes, we observe that the constraints become tighter with increasing spin $a$. 
This trend arises because the frequency modification due to dCS gravity becomes more pronounced as $a$ increases (see Fig.~\ref{fig:omega_1_022}) at a given $\zeta$. 
In particular, for $a \sim 0.7$, the most likely remnant spin of most of the LIGO-Virgo-KAGRA BBHs \cite{LIGO_10, Abbott:2020niy, LIGOScientific:2021djp}, the constraint on $\sqrt{\alpha_{\rm dCS}}$ corresponds to $\zeta = \alpha_{\rm dCS}^2 / M^4 \sim 0.1$, which is within the small-coupling regime for both polar and axial modes.
However, for $a \lesssim 0.4$, the constraint obtained from the polar frequency is significantly weaker than that obtained from the axial frequency. 
This discrepancy occurs because, in the limit $a \rightarrow 0$, polar perturbations decouple from the dCS terms, making it impossible for them to account for the relative measurement uncertainty in this regime. Constraints one would like to place with the polar modes, therefore, exit the regime of validity of the effective-field theory, roughly when $\sqrt{\alpha}_{\rm dCS}>12$ km, and thus, are not valid. 
Overall, Fig.~\ref{fig:constraints} highlights the strong potential of ringdown spectroscopy in obtaining the first meaningful GW-alone constraints on $\sqrt{\alpha_{\rm dCS}}$ in the small-coupling regime.

\begin{table*}[tp!]
\resizebox{\textwidth}{!}{
\begin{tabular}{ccccc}
\hline
Event ~~ & ~~ Median remnant mass ($M_{\odot}$)~~~~ & ~~~~ Median remnant spin ~~ & ~~ axial constraints [km] ~~ & ~~ polar constraints [km] ~~~~~ \\ \hline
GW150914 & 67.4 & 0.67 & 38.2 & 48.1 \\
GW170104 & 48.7 & 0.64 & 35.8 & 46.6 \\
GW170814 & 53.2 & 0.70 & 33.1 & 40.4 \\
GW170823 & 65.4 & 0.72 & 46.9 & 56.0 \\
GW190408$\_$181802 & 41.1 & 0.67 & 27.6 & 34.8 \\
GW190421$\_$213856 & 69.7 & 0.67 & 50.8 & 63.9 \\
GW190503$\_$185404 & 68.6 & 0.66 & 50.4 & 64.2 \\
GW190512$\_$180714 & 34.5 & 0.65 & 26.9 & 34.5 \\
GW190513$\_$205428 & 51.6 & 0.68 & 40.5 & 50.2 \\ 
GW190521 & 156.3 & 0.71 & 102 & 123 \\
GW190521$\_$074359 & 88.0 & 0.71 & 50.5 & 60.3 \\ 
GW190602$\_$175927 & 71 & 0.72 & 52.2 & 62.3 \\ 
GW190708$\_$232457 & 29.5 & 0.69 & 21.7 & 26.8 \\ 
GW190727$\_$060333 & 63.8 & 0.73 & 47.5 & 56.1 \\
GW190828$\_$063405 & 54.9 & 0.75 & 33.3 & 38.6 \\ 
GW190910$\_$112807 & 75.8 & 0.7 & 46.6 & 56.7 \\ 
GW190915$\_$235702 & 57.2 & 0.7 & 40.2 & 49.0 \\ \hline
\end{tabular}
}
\caption{
\label{tab:constraints_LVK}
Projected constraints on $\sqrt{\alpha_{\rm dCS}}$ derived from the axial (middle column) and polar (right column) 022-mode frequencies in dCS gravity, assuming the remnant BH masses and dimensionless spins correspond to the median values estimated for each detected signal \cite{LIGOScientific:2016vlm, Abbott:2017vtc, LIGO_05, LIGO_10, LIGOScientific:2020ibl}.
}
\end{table*}

Let us also estimate the projected constraints using the median remnant mass and dimensionless spin of selected BBH signals detected by the LIGO-Virgo-KAGRA detectors, whose ringdown signals have been analyzed in GR tests by the LIGO-Virgo-KAGRA Collaboration \cite{LIGOScientific:2019fpa, LIGOScientific:2020tif}. 
Some events are excluded from this estimate because their remnant dimensionless spin exceeds 0.75, and thus, our METRICS computations presented here are not valid. 
Despite this exclusion, our estimate still applies to the majority of events detected, as most of them lead to a remnant BH of dimensionless spins $\lesssim 0.75$ \cite{LIGOScientific:2018mvr, LIGOScientific:2020ibl, LIGO_10, Abbott:2020niy}. 
Specifically, we compute $\delta \text{Re}(\omega_{022})$ by dividing the half-width of the 90\% credible interval of the redshifted 022-mode real frequency by the frequency estimate obtained from a full inspiral-merger-ringdown analyses \cite{LIGOScientific:2020tif}.
Table~\ref{tab:constraints_LVK} summarizes the estimated constraints on $\sqrt{\alpha_{\rm dCS}}$, derived using both axial and polar QNM frequencies, based on the remnant mass and spin of the corresponding LVK events.

In the future, next-generation detectors will significantly enhance our ability to constrain modified gravity theories through ringdown signals. 
As demonstrated in \cite{Maselli:2023khq}, if approximately 1000 black-hole ringdown detections are combined, the relative measurement uncertainty in the quasinormal-mode frequencies could be greatly reduced (with the improvement roughly scaling with the inverse of the square root of the number of events). 
This corresponds to roughly a two-order-of-magnitude improvement in frequency precision for a 1000 events, which, in turn, could improve the constraints on the dCS coupling constant by at least a factor of 10, assuming the remnant black holes involved have masses in the range $\sim$ 5-100 $M_{\odot}$.

As we conclude this section, we emphasize that our estimates of the constraints on $\sqrt{\alpha_{\rm dCS}}$ are based on a maximal-research analysis, which does not account for correlations between $\sqrt{\alpha_{\rm dCS}}$ and other signal parameters, particularly the BH mass and spin.
Our estimates also do not fully incorporate the effects of broken isospectrality in dCS gravity, and the numerical uncertainty in METRICS' calculations or the fitting polynomials. 
Our projection relies solely on the dominant, fundamental (022) mode. If additional QNMs are detected with sufficient confidence and high evidence, as suggested in \cite{Isi:2019aib}, the resulting constraints on the dCS coupling parameter should improve.
All these factors influence the resulting constraints.
To properly address these issues, we plan to conduct a comprehensive Bayesian inference analysis of detected ringdown signals, the results of which will be presented in a separate publication.

\section{Concluding Remarks} 
\label{sec:Conclusions}

In this paper, we applied METRICS to compute the gravitational QNM frequencies of rotating BHs in dCS gravity. 
We computed the leading-order modifications to the gravitational QNM frequencies of the $nlm = 022$, 032 and 033 modes of rotating BHs with dimensionless spin parameters $a \lesssim 0.75$. 
The numerical uncertainty of the METRICS frequencies for the 022 mode, which dominates most astrophysical ringdown signals, is $\lesssim 10^{-4}$ for $0 \leq a \leq 0.5$ and $\lesssim 10^{-3}$ for $0.5 \leq a \leq 0.75$. 
Our work is the first accurate computation of gravitational QNM frequencies of rapidly rotating BHs (of $a \sim 0.75$) in dCS gravity. 
Prior to this work, the gravitational QNM frequencies of rotating BHs in dCS gravity had been estimated using slow-rotation expansions, which ought to be valid only up to $a$ values of ${\cal{O}}(0.1)$ \cite{Wagle:2021tam, Srivastava_Chen_Shankaranarayanan_2021}. 
Our work significantly advances the limit of our understanding of BH QNMs in dCS gravity. 

The successful application of METRICS to dCS gravity further demonstrates the power of the METRICS formalism as a unified framework for computing BH QNMs.
DCS gravity is the third gravity theory to which METRICS has been successfully applied, following its earlier applications to GR and EsGB gravity.
The accurate computation of QNM spectra in modified gravity theories using METRICS, along with the development of the modified Teukolsky formalism~\cite{Li:2022pcy, Wagle:2023fwl, Li:2023ulk, Cano:2023tmv, Hussain:2022ins}, suggests that there are no fundamental obstacles to computing the QNM spectra in modified gravity theories, provided that the corresponding field equations and the metric of rotating BHs are known.
Despite these successes, the QNM spectra of \textit{extremal} BHs in modified gravity remain largely unexplored, primarily due to the lack of sufficiently accurate solutions for the background metric and scalar field in modified gravity theories.
To address this, we plan to develop techniques for the accurate construction of extremal BH backgrounds in modified gravity theories in the future. 

The METRICS frequencies in dCS gravity offer several potential applications. 
In the context of data analysis, these frequencies, along with their optimal fitting expressions, can be used to analyze observed ringdown signals, providing a means to test dCS gravity. 
When combined with inspiral tests, the results from ringdown tests could significantly enhance GW constraints on dCS gravity. 
Additionally, from a theoretical perspective, the METRICS frequencies can aid in understanding numerical simulations of dCS gravity by enabling precise mode identification, contributing to a deeper understanding of this gravity theory.

\section*{Acknowledgements}

The authors acknowledge the support from the Simons Foundation through Award No. 896696, the NSF through award PHY-2207650, and NASA through Grant No. 80NSSC22K0806. 
The authors would like to thank Pablo Cano, Simon Maenaut, Andrea Maseli, Dongjun Li, and Pratik Wagle for insightful discussion about this work. 
We are also indebted to Dongjun Li for providing $\omega^{(1)} (\rm MTE)$ used in this paper, which allowed us to determine the accuracy of our fitting algorithm. 
The calculations and results reported in this paper were produced using the computational resources of the Illinois Campus Cluster, a computing resource that is operated by the Illinois Campus Cluster Program (ICCP) in conjunction with National Center for Supercomputing Applications (NCSA), and is supported by funds from the University of Illinois at Urbana-Champaign, and used Delta at NCSA through allocation PHY240142 from the Advanced Cyberinfrastructure Coordination Ecosystem: Services $\&$ Support (ACCESS) program, which is supported by National Science Foundation Grants No. 2138259, No. 2138286, No. 2138307, No. 2137603, and No. 2138296.
The author would like to specially thank the investors of the IlliniComputes initiatives and GravityTheory computational nodes for permitting the authors to execute runs related to this work using the relevant computational resources. 

\section*{Data Availability}

The data that support the findings of this article are not publicly available. The data are available from the authors upon reasonable request.

\appendix

\section{Additional Tables}
\label{sec:Appendix_B}

The following tables present additional details related to the results described in the main body of this paper. 
In particular, Tables~\ref{tab:omega_1_022}, \ref{tab:omega_1_033}, and \ref{tab:omega_1_021} present the dCS corrections to the quasinormal frequencies for the $022$, $033$ and $032$ modes respectively, for various choices of BH spin. 
The dCS shift of the frequencies is estimated using the results a $\zeta = 5 \times 10^{-5}$ and $\zeta = 10^{-4}$. 
Meanwhile, Table~\ref{tab:omega_1_022_scalar} presents the sGB quasinormal frequencies of the scalar mode. Tables~\ref{tab:poly_fit_coeffs} and~\ref{tab:poly_fit_coeffs_uncertainty} show the coefficients of the fitting polynomials, as well as their uncertainties, respectively.   

\begin{figure*}[htp!]
\centering  
\subfloat{\includegraphics[width=0.47\linewidth]{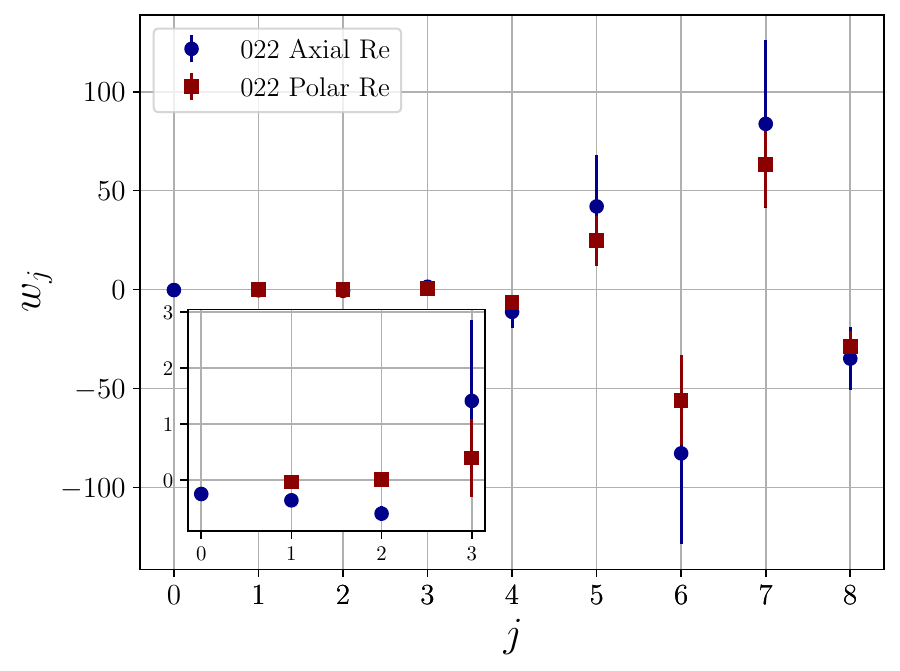}}
\subfloat{\includegraphics[width=0.47\linewidth]{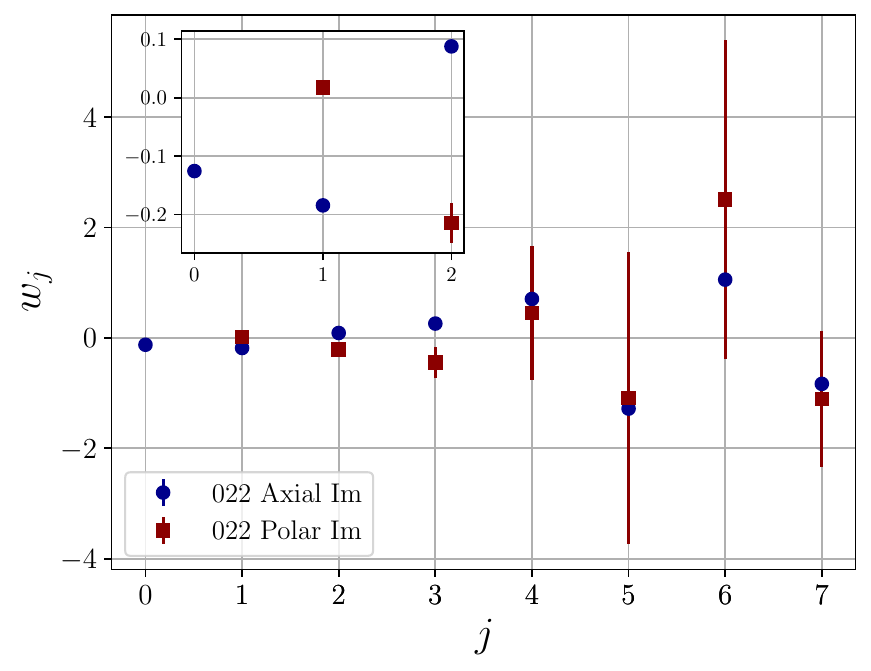}}
\caption{Best-fit values (scatters) and uncertainties (error bars) of the fitting polynomial coefficients ($w_j$) of the 022 axial (dark blue) and polar (dark red) modes. 
Since the magnitude of the best-fit value and uncertainty of the first few coefficients is significantly smaller than that of the subsequent coefficients, the information of the first few coefficients are visualized in an inset panel. 
We observe that the magnitudes of the best-fit values and uncertainties grow significantly with $j$.
This is because the coefficients of the last few terms of the fitting polynomials are strongly correlated. 
}
\label{fig:w_j}
\end{figure*}

\begin{figure*}[htp!]
\centering  
\subfloat{\includegraphics[width=0.47\linewidth]{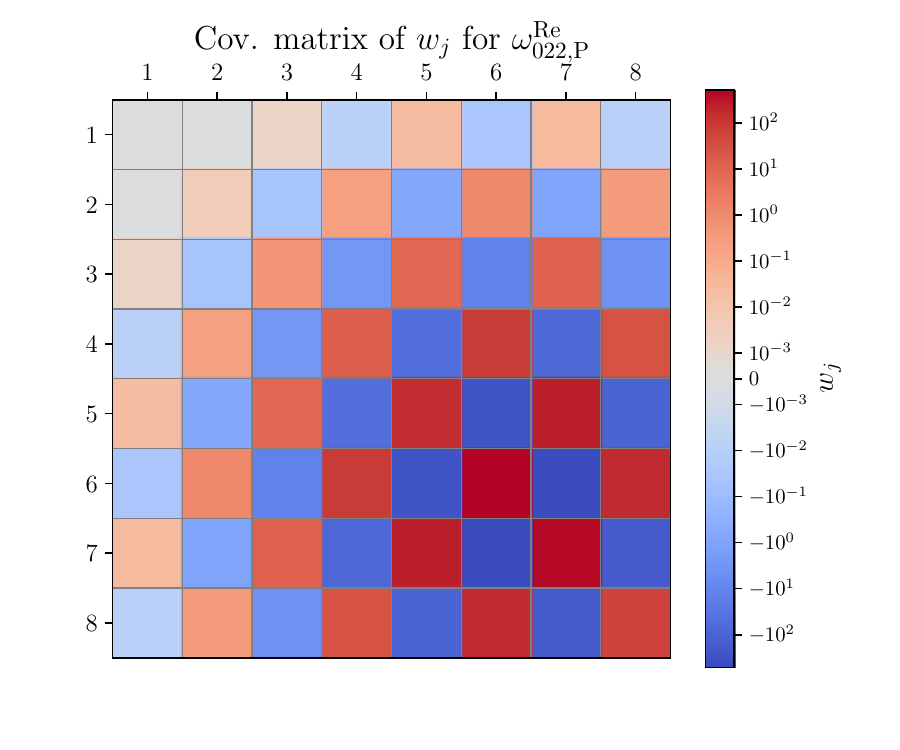}}
\subfloat{\includegraphics[width=0.47\linewidth]{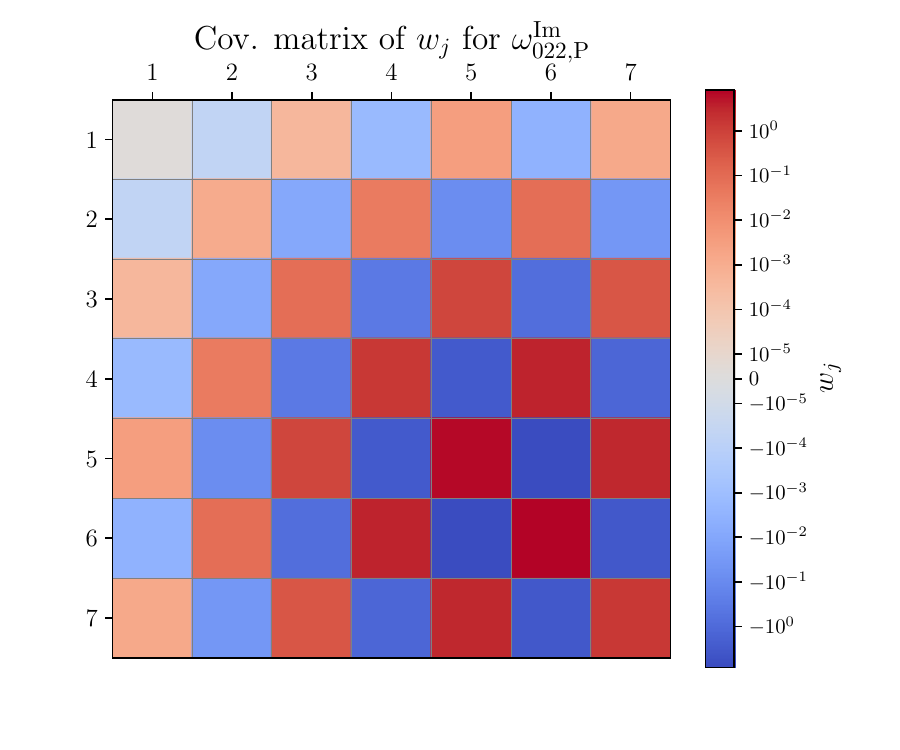}}
\caption{Color plot of the covariance matrix of $w_j$ of the fitting polynomials of the real (left) and imaginary (right) parts of the 022 polar mode frequency. 
Blue hues represent negative values, and red hues represent positive values. 
Observe that the coefficients of the last few terms of the fitting polynomials are strongly correlated, indicating that an alternative fitting expression might better fit the METRICS dCS frequencies. 
}
\label{fig:Cov_matrix}
\end{figure*}

An important result of this paper is the fitting polynomials that can rapidly and accurately compute the dCS QNM spectra. 
The best-fitted value and uncertainty of the coefficients of the polynomials are respectively listed on Tables~\ref{tab:poly_fit_coeffs} and~\ref{tab:poly_fit_coeffs_uncertainty}. 
We plot the coefficients of the fitting polynomial of the 022 mode frequencies, as examples, on Fig.~\ref{fig:w_j}, with numerical uncertainties shown with error bars.
Observe that the magnitude of the best-fit value and uncertainty of $w_j$ increase rapidly with $j$. 
This is a general feature of the fitting polynomials that we have observed also when fitting QNM spectra of other modes. 
To further understand this feature, we show the covariance matrix of $w_j$ of the 022 polar mode in Fig.~\ref{fig:Cov_matrix}. 
Observe that the magnitude of the off-diagonal elements of the last few $j$ is large, indicating that the coefficients of the last few terms are strongly correlated. 
This indicates that an alternative fitting expression might better fit the METRICS frequencies. 
However, as shown in Fig.~\ref{fig:fitting_polynomials_check}, the relative fractional error of our fitting polynomials is $< 10^{-2}$. 
Thus, our polynomial fit at the best-fit values can still accurately compute the METRICS frequencies. 
We leave the exploration of other possible fitting expressions to future work. 

\begin{figure}[htp!]
\centering  
\subfloat{\includegraphics[width=\columnwidth]{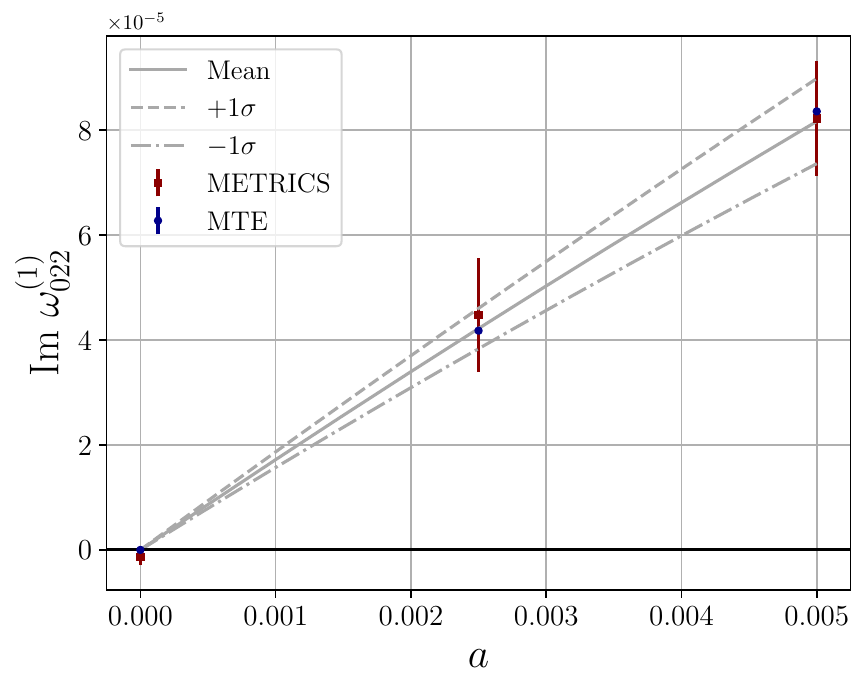}}
\caption{
Comparison of the imaginary part of $\omega_{022, P}^{(1)}$ computed using METRICS (red squares), the modified Teukolsky formalism (blue circles), and our fitting expression (solid gray line). 
The fitting expression is truncated at second order in $ a $, with $ w_0 $ and $ w_1 $ at their best-fit values. 
The dashed and dash-dotted gray lines represent the uncertainty range, corresponding to $ w_0 $ and $ w_1 $ shifted by $ \pm 1\sigma $. 
Both METRICS and modified Teukolsky results fall within this region, confirming the consistency of the three approaches. 
The observed discrepancy in $ w_1 $ is attributed to fitting artifacts rather than a fundamental inconsistency.
}
 \label{fig:MTE_check_fitting}
\end{figure}

By comparing the values of $w_0$ and $w_1$ listed in Table \ref{tab:poly_fit_coeffs} with the results listed in Tables I and II in \cite{Li:2025fci} (after appropriate rescaling to reconcile differences between our definitions of the coupling constant), we find that our results agree well, despite a larger difference in the imaginary part of $w_1$ of the polar modes. 
To understand the difference, Fig.~\ref{fig:MTE_check_fitting} compares the imaginary part of the $022$ frequency, $\text{Im} (\omega_{022, P}^{(1)})$, computed using METRICS (red squares), calculated from our fitting expression truncated at second order in $a$ (gray lines), and that obtained from the modified Teukolsky formalism (blue circles). 
Observe that the METRICS and modified Teukolsky results are very close to each other, with a slightly larger difference at spin of $0.0025$.
The dashed (dashed-dotted) gray line is the same as the solid gray line, but with $w_0$ and $w_1$ the best-fit value plus (minus) one $\sigma (w_j)$. 
The region enclosed by the dashed and the dashed-dotted lines, thus, represents the value of $\text{Im} (\omega_{022, P}^{(1)})$ that is consistent within the uncertainty of the fitting expression. 
Observe that both the METRICS and modified-Teukolsky frequencies are within this region, indicating the consistency between the calculations. 


\begin{table*}[htb]
\resizebox{\textwidth}{!}{
\begin{tabular}{c|c|c|c|c|c|c|c}
\hline
$a$ & $N_{\rm a}$ & $\omega^{(1)}_{\rm (A)}$ & $10^4 \times \delta^{\rm (A)}$ & $\omega^{(1)}_{\rm (P)}$ & $10^4 \times \delta^{\rm (P)}$ \\ \hline
0 & 0 & $ -0.2460405768551833 -0.12548170333067551i$ & $ 0.0381$ & $ 2.8283877541067245 \times 10^{-7} -1.3080838730760558\times 10^{-6}i $ & $ 0.0158 $ \\
0.005 & 2 & $ -0.24785316562790285 -0.12640118239992554i $ & $ 0.176$ & $ -0.00015481199751754047 + 8.221472002378407\times 10^{-5} i $ & $ 0.109$ \\
0.1 & 4  & $  -0.2872701816159066-0.14270190229282534i $ & $ 0.317$ & $ -0.003009053170057409 -0.0008137430424489493i $ & $ 0.149 $ \\
0.2 & 8  & $ -0.3392802984458994-0.15596769814808242i$ & $ 0.322$ & $ -0.008004096259119728-0.008156589024320959i $ & $ 0.159$ \\
0.3 & 10  & $ -0.40270887230336005-0.1626898826762333i$ & $ 0.532$ & $ -0.01993915233055915-0.023582419332679006i $ & $ 0.300$ \\
0.4 & 14  & $ -0.477668564719445-0.16075985408127055i$ & $ 0.992$ & $ -0.046485819104627464-0.04721541430349899i $ & $ 0.457$ \\
0.5 & 16  & $ -0.5649437171180661-0.14947432402040778i$ & $ 1.73 $  & $ -0.09896521597063425-0.07629887294351685i $ & $ 1.57 $ \\
0.6 & 22  & $ -0.6693309844045422-0.1312780966998435i $ & $ 2.69$  & $ -0.1948634764259947-0.10428706507154202i $ & $ 2.50$ \\
0.690 & 30  & $ -0.7890154828547377-0.1152836155547013i$ & $ 4.41$  & $ -0.3429790248344785-0.11790979732397841i $ & $ 3.50 $ \\
0.753 & 36  & $ -0.9080224040028065-0.11054335780380233i$ & $ 5.19 $  & $ -0.5114075139526467-0.1146417076239047i$ & $ 5.69 $ \\ \hline
\end{tabular}
}
\caption{
\label{tab:omega_1_022}
$\omega^{(1)}$ of the $nlm=022$-mode gravitational perturbations of rotating BHs in dynamical Chern-Simons (dCS) gravity at different dimensionless spins $a$ (first column). 
The superscripts (A) and (P) respectively stand for axial and polar perturbations. 
$\delta$ is the numerical uncertainty of the frequency calculations, which is the respective minimal backward modulus difference of the real and imaginary parts of the frequency. 
The numerical value of the real and imaginary parts of $\omega^{(1)}$ is rounded to the nearest decade which is larger than the numerical uncertainty. 
}
\end{table*}

\begin{table*}[htb]
\resizebox{\textwidth}{!}{
\begin{tabular}{c|c|c|c|c|c}
\hline
$a$ & $N_{\rm a}$ & $\omega^{(1)}_{\rm (A)}$ & $10^4 \times \delta^{\rm (A)}$ & $\omega^{(1)}_{\rm (P)}$ & $10^4 \times \delta^{\rm (P)}$ \\ \hline
0 & 0 & $ -0.9127523524803032 - 0.1646154968058758 i $ & $ 0.00174 $ & $ -1.3322726077828663\times 10^{-12} - 9.43685836549922\times 10^{-13} i $ & $ 0.00101 $ \\
0.005 & 2 & $ -0.9182584646111505 - 0.1650679401681875 i$ & $ 0.0224 $ & $ -0.0004526322801288097 - 7.976573525490674\times 10^{-6} i $ & $ 0.00610 $ \\
0.1 & 4  & $ -1.0334533453414616 - 0.17187830398362594 i $ & $ 0.202 $ & $ -0.009793649200446218 - 0.0018714149254726604i $ & $ 0.0170 $ \\
0.2 & 8  & $ -1.1771518016306726 - 0.17314794895245117 i $ & $ 0.297 $ & $ -0.02302540236359687 - 0.008667312013599156i $ & $ 0.0524 $ \\
0.3 & 10  & $ -1.3485252334294382 - 0.16540758252862192 i $ & $ 0.526 $ & $ -0.04458242169445468 - 0.022500573939442735i $ & $ 0.117 $ \\
0.4 & 14  & $ -1.5533633232305288 - 0.14526451608198568 i $ & $ 0.878 $ & $ -0.0834938364060477 - 0.045014699193518214i $ & $ 0.340 $ \\
0.5 & 16  & $ -1.7991571796602903 - 0.10971870829814231 i $ & $ 1.10 $  & $ -0.15625316120618374 - 0.07593603435668841i $ & $ 0.810 $ \\
0.6 & 22  & $ -2.096018808055661 - 0.05795781095226621 i $ & $ 2.79 $  & $ -0.2935664161460019 - 0.11093430483776007i $ & $ 1.64 $ \\
0.690 & 30  & $ -2.4260075161943915 - 0.0039667500843716215 i $ & $ 3.98 $  & $ -0.5216562726294325 - 0.13412047594566778i $ & $ 7.29 $ \\
0.753 & 36  & $ -2.7184602708238432 + 0.028960259121882555 i $ & $ 14.3 $  & $ -0.7977553784543139 - 0.13555287807781816i $ & $ 9.31 $ \\ \hline
\end{tabular}
}
\caption{
\label{tab:omega_1_033}
Identical to Table ~\ref{tab:omega_1_022}, except that $nlm = 033$. 
}
\end{table*}

\begin{table*}[htp!]
\resizebox{\textwidth}{!}{
\begin{tabular}{c|c|c|c|c|c|c|c}
\hline
$a$ & $N_{\rm a}$ & $\omega^{(1)}_{\rm (A)}$ & $10^4 \times \delta^{\rm (A)}$ & $\omega^{(1)}_{\rm (P)}$ & $10^4 \times \delta^{\rm (P)}$ \\ \hline
0 & 0 & $-0.9127264243454835i -0.16460569809981143i$ & $0.00170$ & $ -2.2384202324213632 \times 10^{-8} + 9.394643177296176\times 10^{-9}$ & $ 0.000515$ \\
0.005 & 2 & $-0.9163526463407625-0.16475504565823482i$ & $  0.0838$ & $ 5.9403428925804746\times 10^{-5} + 9.577803218928367\times 10^{-7}$ & $ 0.00682 $ \\
0.1 & 4  & $-0.9896438921456346-0.16832205933858435i$ & $ 0.0738$ & $-0.005099968769288549 -0.0013248844867635922i$ & $0.0313$ \\
0.2 & 8  & $-1.0728614469794284-0.16571567413668417i$ & $ 0.0626$ & $-0.008304566486721388 -0.0061843935221039615i$ & $ 0.0358$ \\
0.3 & 10  & $-1.1618053852456125-0.15507569539933522i$ & $ 0.146$ & $-0.009692024970172205 -0.01624016538046188i$ & $ 0.135$ \\
0.4 & 14  & $-1.2541391182462502-0.13456732388577505i$ & $ 0.796$ & $ -0.00937489645247046  -0.03325733062147201i$ & $ 0.416$ \\
0.5 & 16  & $-1.3488110435762155-0.1037224787497714i$ & $ 1.65$  & $ -0.007172837895150912 -0.05917126234911792i$ & $ 1.02$ \\
0.6 & 22  & $-1.438996920239855-0.06315351771852007i$ & $ 4.78$  & $ -0.0019059593510296288 -0.09671654772643674i$ & $  2.32$ \\
0.690 & 30  & $-1.5134658174484594-0.02512170418959085i$ & $ 9.06$  & $ 0.007600477208230129 -0.1422044740407492i$ & $ 5.07 $ \\
0.753 & 36  & $-1.5624886645910172-0.003304518601245734i$ & $8.57   $  & $ 0.019597337286770076  -0.18601522986843694i$ & $ 7.59$ \\ \hline
\end{tabular}
}
\caption{
\label{tab:omega_1_021}
Identical to Table ~\ref{tab:omega_1_022}, except that $nlm = 032$. 
}
\end{table*}

\begin{table*}[htp]
\resizebox{\textwidth}{!}{
\begin{tabular}{c|cc|cc|cc|}
\hline
\multirow{2}{*}{$w_j$} & \multicolumn{2}{c|}{022}                                                     & \multicolumn{2}{c|}{033}                                                 & \multicolumn{2}{c|}{032}                                                   \\ \cline{2-7} 
                       & \multicolumn{1}{c|}{(A)}                         & (P)                       & \multicolumn{1}{c|}{(A)}                      & (P)                      & \multicolumn{1}{c|}{(A)}                      & (P)                        \\ \hline
$w_{0}$                  & \multicolumn{1}{c|}{$-0.246041 -0.125482i$}      & $-$     & \multicolumn{1}{c|}{$-0.912752-0.164615 i$}   & $ - $    & \multicolumn{1}{c|}{$ -0.912726 -0.164606i $}    & $ - $     \\ 
$w_{1}$                  & \multicolumn{1}{c|}{$-0.35908 -0.184144 i$}      & $-0.0307839 + 0.017411i $     & \multicolumn{1}{c|}{$-1.09512-0.0911042i$}   & $ -0.0901535 -0.000837525 i$    & \multicolumn{1}{c|}{$  -0.721554 -0.0266668i $}    & $ 0.0213119 + 0.000953349i $     \\ 
$w_{2}$                  & \multicolumn{1}{c|}{$-0.595672 + 0.0876598 i$}      & $0.013835 -0.214353 
 i $     & \multicolumn{1}{c|}{$-1.2327+0.116646i$}   & $ -0.0763526 -0.149877 i$    & \multicolumn{1}{c|}{$-0.815162 -0.717186i $}    & $ -1.98616 -0.154152 i$     \\ 
$w_{3}$                  & \multicolumn{1}{c|}{$1.41373 + 0.258192i$}      & $0.398761 -0.447376 i $     & \multicolumn{1}{c|}{$2.37515+0.763688i$}   & $  0.452064 -0.248466 i$    & \multicolumn{1}{c|}{$ 6.14597 + 9.84411i $}    & $ 20.3576 + 0.42749i$     \\ 
$w_{4}$                  & \multicolumn{1}{c|}{$-11.2668 + 0.703524i$}      & $-6.66087 +0.448166 i$     & \multicolumn{1}{c|}{$-16.7427-0.927771i$}   & $ -6.70664 -0.508568 i$    & \multicolumn{1}{c|}{$  -38.4939 -49.307i $}    & $ -101.368 -4.25973 i$     \\ 
$w_{5}$                  & \multicolumn{1}{c|}{$42.008 -1.28144i$}      & $24.7362 -1.091 i $     & \multicolumn{1}{c|}{$52.5812+1.12866i$}   & $ 25.7745 + 1.18438 i$    & \multicolumn{1}{c|}{$ 131.55 + 142.024i $}    & $ 281.599 + 14.4685i$     \\ 
$w_{6}$                  & \multicolumn{1}{c|}{$-82.8112 + 1.05377 i$}      & $-55.9612 + 2.50703i $     & \multicolumn{1}{c|}{$-97.6497+0.865323i$}   & $ -60.5245 -1.60982i$    & \multicolumn{1}{c|}{$ -244.639-230.178i $}    & $  -441.536 -27.5577 i $     \\ 
$w_{7}$                  & \multicolumn{1}{c|}{$83.7508-0.833429i$}      & $63.124  - 1.10815i$     & \multicolumn{1}{c|}{$97.2953 -1.91376 i$}   & $ 69.0726 + 1.64983i $    & \multicolumn{1}{c|}{$ 233.961 + 195.929 i $}    & $ 364.751 + 27.2273 i $     \\ 
$w_{8}$                  & \multicolumn{1}{c|}{$-34.8784 $}      & $-28.7982 $     & \multicolumn{1}{c|}{$-40.6093$}   & $-32.9995  $    & \multicolumn{1}{c|}{$-89.5753-68.4985i$}    & $ -123.1 -10.9198i $     \\ \hline
\end{tabular}
}
\caption{
\label{tab:poly_fit_coeffs}
The coefficients $w_j$ of the optimal fitting polynomial (c.f. Eq.~\eqref{eq:omega_1_fitted}) to the axial and polar frequencies of the $nlm=022, 033$ and 032. 
Note that the degree of the optimal fitting polynomial of the real and imaginary parts of the same parity of a given mode can be different. }
\end{table*}

\begin{table*}[htp]
\resizebox{\textwidth}{!}{
\begin{tabular}{c|cc|cc|cc|}
\hline
\multirow{2}{*}{$\sigma(w_j)$} & \multicolumn{2}{c|}{022}                                                     & \multicolumn{2}{c|}{033}                                                 & \multicolumn{2}{c|}{032}                                                   \\ \cline{2-7} 
                       & \multicolumn{1}{c|}{(A)}                         & (P)                       & \multicolumn{1}{c|}{(A)}                      & (P)                      & \multicolumn{1}{c|}{(A)}                      & (P)                        \\ \hline
$\sigma_{w_0}$                  & \multicolumn{1}{c|}{$ 4.80\times10^{-6} + 4.22\times10^{-7} i $} & $ - $  & \multicolumn{1}{c|}{$ 7.29 \times 10^{-8} + 8.18\times 10^{-8} i$} & $ - $ & \multicolumn{1}{c|}{$5.67\times10^{-7} + 3.43\times10^{-7} i$} & {$-$}    \\
$\sigma_{w_1}$                  & \multicolumn{1}{c|}{$0.00513 + 0.000404i $}     & $ 0.00244 + 0.00145i $ & \multicolumn{1}{c|}{$ 0.000221 + 0.000239i$} & $ 0.000245 + 0.0000810i$  & \multicolumn{1}{c|}{ $0.00637 + 0.00385 i$} & {$0.000416 + 0.000160i $} \\
$\sigma_{w_2}$                  & \multicolumn{1}{c|}{$0.135 + 0.00915i $}     & $ 0.0644 + 0.0335 i$  & \multicolumn{1}{c|}{$ 0.0131 + 0.00837i $}  & $ 0.0177 + 0.00251i$ & \multicolumn{1}{c|}{$0.173 + 0.104i$} & {$0.0284 + 0.0109 i$}    \\
$\sigma_{w_3}$                  & \multicolumn{1}{c|}{$1.45 + 0.0777i $}     & $ 0.692 + 0.288 i$   & \multicolumn{1}{c|}{$0.235 + 0.105i $}   & $ 0.363 + 0.0343 i$   & \multicolumn{1}{c|}{$2.03 + 1.23 i$} &{$0.572 + 0.220 i$}       \\
$\sigma_{w_4}$                  & \multicolumn{1}{c|}{$8.17+0.321i $}       & $ 3.93 + 1.21i $      & \multicolumn{1}{c|}{$1.75 + 0.566i$}    & $ 2.98 + 0.211 i$   & \multicolumn{1}{c|}{$13.1 + 7.90i $} &{$4.67 + 1.79 i$}       \\
$\sigma_{w_5}$                  & \multicolumn{1}{c|}{$25.9+0.687i $}      & $ 12.6 + 2.65 i$     & \multicolumn{1}{c|}{$ 6.47 + 1.47 i$}   & $ 12.0 + 0.616i $     & \multicolumn{1}{c|}{$ 47.3 + 28.6i $} & {$ 18.7 + 7.18 i$}       \\
$\sigma_{w_6}$                  & \multicolumn{1}{c|}{$45.8+0.731i $}      & $ 22.7 + 2.88i$      & \multicolumn{1}{c|}{$ 12.6 + 1.80i$}     & $ 25.3 + 0.836i $     & \multicolumn{1}{c|}{$ 94.2 + 56.9 i $} & {$39.1 + 15.0 i$}    \\
$\sigma_{w_7}$                  & \multicolumn{1}{c|}{$42.3+0.305i$}      & $ 21.3 + 1.23 i $     & \multicolumn{1}{c|}{$12.4 + 0.844i $}   & $ 26.6 + 0.423i $    & \multicolumn{1}{c|}{$ 95.4 + 57.7 i $} &  {$ 40.6 + 15.6 i$}      \\ 
$\sigma_{w_8}$                  & \multicolumn{1}{c|}{$15.7$}      & $ 8.06 $     & \multicolumn{1}{c|}{$4.82$}   & $11.0$    & \multicolumn{1}{c|}{$ 38.3 + 23.2i $} &  {$ 16.6 + 6.36i $}    \\ \hline
\end{tabular}
}
\caption{
\label{tab:poly_fit_coeffs_uncertainty}
Uncertainty of the real and imaginary parts of $w_j$ of the fitting polynomial (c.f. Eq.~\eqref{eq:omega_1_fitted}) to the axial and polar frequencies of the $nlm=022, 033$ and 032.}
\end{table*}

\begin{table*}[htp]
\begin{tabular}{c|c|c|c}
\hline
$a$ & $N_{\rm a}$ & $\omega^{(1)}_{\rm (S)}$ & $\delta^{\rm (S)}$ \\ \hline
0.005 & 2 &  $0.5825756141022715 + 0.07864087230312451i$ & $(2.17+3.71i) \times 10^{-5}$  \\
0.1 & 4 &  $0.6163527093170601 + 0.05205163863673137i$ & $(1.95+3.71i) \times 10^{-5}$ \\
0.2 & 8 &  $0.6239254817008097 + 0.03539103326303573i$ & $(2.32+7.11i) \times 10^{-6}$ \\
0.3 & 10 &  $0.6387675729171479 - 0.002334530413770126i $ & $(2.56+5.30i) \times 10^{-6}$ \\
0.4 & 14 &  $0.621999315383891 + 0.0588440654914917i$ & $0.000516 + 0.0409 i $ \\
0.5 & 16 &  $0.6018143322164952 + 0.08718647592308519i$ & $0.0182 + 0.0156 i$ \\
0.6 & 22 &  $0.5185366269559957 + 0.061717972931849975i $ & $0.0430 + 0.0322 i $ \\
0.69 & 30 &  $0.5612749413287468 + 0.17862312362577298i$ & $ 0.518 + 0.0642 i $ \\
0.753 & 46 &  $0.0.4874755081242025 + 0.24583389558182145i$ & $ 0.0675 + 0.0195 i$ \\ \hline
\end{tabular}
\caption{
\label{tab:omega_1_022_scalar} 
Identical to Table ~\ref{tab:omega_1_022}, except that the mode is the scalar mode. 
}
\end{table*}

\bibliography{ref}

\end{document}